\documentclass[10pt,journal,compsoc]{IEEEtran}

\usepackage[T1]{fontenc}
\ifCLASSOPTIONcompsoc

\usepackage{cite}

\usepackage{bm}
\usepackage{amsmath}
\usepackage{amsthm}
\usepackage{float}
\usepackage{setspace}
\usepackage{amssymb}
\usepackage{stfloats}
\usepackage{cite}
\usepackage{ragged2e}
\usepackage[ruled,vlined,linesnumbered]{algorithm2e}
\usepackage{amsfonts}
\usepackage{mathrsfs}
\usepackage{amsmath,amsthm}
\usepackage{array,booktabs}
\usepackage{subfigure}
\usepackage{multirow}
\usepackage{cuted}
\usepackage{multicol}
\usepackage{graphicx}
\usepackage{subfigure}
\usepackage{graphicx,xcolor,bm}
\usepackage{hyperref}
\usepackage{threeparttable}
\usepackage{dcolumn}
\usepackage{setspace}
\usepackage{makecell}
\usepackage{lipsum}
\usepackage{enumerate}
\usepackage{mathrsfs}
\usepackage{bbm}
\usepackage{amsmath}
\usepackage{array}
\usepackage[bottom]{footmisc}

\usepackage{cite,bm}
\graphicspath{{figures/}}
\begin{document}
	\title{Oh-Trust: Overbooking and Hybrid Trading-empowered Resource Scheduling with Smart Reputation Update over \\Dynamic Edge Networks}
	
\author{Houyi Qi, Minghui Liwang, \IEEEmembership{Senior Member}, \IEEEmembership{IEEE},
					Liqun Fu, \IEEEmembership{Senior Member}, \IEEEmembership{IEEE},
					Xianbin Wang, \IEEEmembership{Fellow}, \IEEEmembership{IEEE},
					Huaiyu Dai, \IEEEmembership{Fellow}, \IEEEmembership{IEEE},
					Xiaoyu Xia, \IEEEmembership{Senior Member}, \IEEEmembership{IEEE}
	
	\thanks{
 This work was supported in part by the National Natural Science Foundation of China under Grant no. 62271424; Shanghai Pujiang Programme under Grant no. 24PJD117, the Shanghai Municipal Science and Technology Major Project under Grant no. 2021SHZDZX0100; the Chinese Academy of Engineering, Strategic Research and Consulting Program under Grant no. 2023-XZ-65. H. Qi (houyiqi@tongji.edu.cn) and M. Liwang (minghuiliwang@tongji.edu.cn) are with the Shanghai Research Institute for Intelligent Autonomous Systems, the State Key Laboratory of Autonomous Intelligent Unmanned Systems, Department of Control Science and Engineering, Tongji University, Shanghai, China. L. Fu (liqun@xmu.edu.cn) is with the School of Informatics, Xiamen University, Fujian, China. X. Wang (xianbin.wang@uwo.ca) is with the Department of Electrical and Computer Engineering, Western University, Ontario, Canada. H. Dai (hdai@ncsu.edu.) is with the Department of Electrical and Computer Engineering, North Carolina State University, NC, USA. X. Xia (xiaoyu.xia@rmit.edu.au) is with the School of Computing Technologies, RMIT University, Melbourne, Victoria, Australia. Corresponding author: Minghui Liwang
		
	}}

	\IEEEtitleabstractindextext{
		\begin{abstract}
			\justifying
Incentive-driven computing resource sharing is crucial for meeting the ever-growing demands of emerging mobile applications. Although conventional spot trading offers a solution, it frequently leads to excessive overhead due to the need for real-time trading related interactions. Likewise, traditional futures trading, which depends on historical data, is susceptible to risks from network dynamics. This paper explores a dynamic and uncertain edge network comprising a computing platform, e.g., an edge server, that offers computing services as resource seller, and various types of mobile users with diverse resource demands as buyers, including fixed buyers (FBs) and uncertain occasional buyers (OBs) with fluctuating needs. To facilitate efficient and timely computing services, we propose an \underline{o}verbooking- and \underline{h}ybrid \underline{tr}ading-empowered reso\underline{u}rce \underline{s}cheduling mechanism with repu\underline{t}ation update, termed Oh-Trust. Particularly, our Oh-Trust incentivizes FBs to enter futures trading by signing long-term contracts with the seller, while simultaneously attracting OBs to spot trading, enhancing resource utilization and profitability for both parties. Crucially, to adapt to market fluctuations, a smart reputation updating mechanism is integrated, allowing for the timely renewal of long-term contracts to optimize trading performance. Extensive simulations using real-world datasets demonstrate the effectiveness of Oh-Trust across multiple evaluation metrics.
		\end{abstract}
		
		\vspace{-1mm}
}
	
\maketitle
\IEEEdisplaynontitleabstractindextext

\IEEEpeerreviewmaketitle

\section{Introduction}
The continuous evolution of next-generation information and artificial intelligence (AI) technologies has witnessed an ever-growing global AI market expected to expand from \$64 billion in 2022 to nearly \$251 billion by 2027. This gives rise to numerous computation-intensive applications, e.g., the metaverse and large language models like ChatGPT and Sora~\cite{chatgpt,survey 1}, where deploying AI-based applications directly on devices' side has emerged as a prominent future trend~\cite{AI1,AI21}. Despite this paradigm shift, the core of these applications lies in their ability to swiftly and accurately analyze vast amounts of data using intelligent algorithms. These processes demand substantial computational power, posing significant challenges for mobile users (MUs) with limited resources \cite{survey 2,edge, edge_1, survey 3,survey 4,survey 5,survey 6}. To alleviate their heavy workloads, mobile edge computing is emerging as a viable solution, offering responsive and cost-effective computing services close to the network edge \cite{survey 2, survey 3,edge,edge_1}, which, however, may still encounter resource shortage when facing sudden and high resource demand. To efficiently manage limited resources while attracting more MUs for resource sharing, designing appropriate incentives becomes both urgent and essential\cite{trading,price,price1}.
{Nowadays, mainstream cloud and edge service providers, such as Amazon Web Services and Google Cloud Platform, commonly employ market-oriented strategies, including fixed-price provisioning, spot instances, and reserved contracts, to flexibly allocate computing resources and balance supply–demand fluctuations \cite{GCP}. These practices have demonstrated that market-based mechanisms can significantly improve utilization and mitigate risks of over- or under-provisioning in large-scale uncertain environments.} {In particular, in a trading, the computing platform functions as a \textit{seller} with paid services, while MUs with computation-intensive tasks act as \textit{buyers}, paying for accessing resources \cite{survey 4,Related spot 1,Related spot 2}.}

\subsection{Motivation}
We conduct our motivations by answering the following crucial questions:
\textit{Q1. Why spot trading is not enough to support resource sharing in dynamic networks?} One most popular resource trading mode refers to \textit{spot trading}\cite{Related spot 1,Related spot 2,Related spot 3}, where transactions are based on real-time agreements between sellers and buyers, determined by the prevailing market and network conditions. \cite{Related futures 3, survey 5}. Although spot trading offers an intuitive procedure, it generally rely on onsite decision-making with significant drawbacks, including excessive overhead~\cite{Related futures 1,Related futures 2} and potential trading failures~\cite{Related futures 3}, especially in dynamic and uncertain network environments. 
{For instance, onsite decision-making, such as negotiating resource quantities and prices, can be both time- and energy-intensive.} {In decentralized spot trading, multiple rounds of bid–feedback–rebid interactions incur significant communication and computation overhead, reducing service efficiency.} {Furthermore, the limited and fluctuating availability of resources may prevent some buyers from obtaining needed services, even after substantial negotiation efforts \cite{Related futures 1,Related futures 3}.} To this end, researchers start to think about if trading agreements (also called contracts) can be pre-signed ahead of practical transactions, thus facilitating a \textit{futures trading} mode.

In general, futures trading allows participants to establish long-term contracts for future needs\cite{Related futures 1,Related futures 2,Related futures 3}. With these contracts in place, contractual participants can engage in trading with full confidence that the agreed-upon terms will be upheld, significantly reducing the overhead associated with decision-making during practical transactions. Nevertheless, we opt to answer the second question: \textit{Q2. What challenges can be brought by future trading?} Since agreements are pre-negotiated, relying heavily on the evaluation of historical information, many risks can be incurred. For instance, improper contract terms (e.g., unreasonable prices, improper quantity of trading resources) can leave negative impacts on utilities of participants, even leading to failures in service delivery, especially when statistical analysis cannot accurately predict market conditions, e.g., future resource supply and demand. Thus, we move to the next question \textit{Q3. How overbooking can cope with the dynamics of resource supply and demand?} Since futures trading allows for the presale of resources (e.g., enabling resources to be booked before actual demand arises), traditional booking strategies (where the amount of booked resources cannot exceed the theoretical supply) face difficulties to handle ever-changing resource demand and supply~\cite{Related futures 3}. To cope with these challenges, various commercial sectors, such as airlines\cite{Overbook1}, hotels\cite{Overbook2}, and telecom companies\cite{Overbook3}, have adopted the practice of overbooking. This strategy allows the booked resources (e.g., flight tickets, hotel rooms, and spectrum) to exceed their practical supply, bringing its effectiveness in enhancing system profits and resource utilization amidst dynamics\footnote{Overbooking allows the seller to sign more contracts than its actual capacity, e.g., an edge server that can handle at most 3 tasks may still sign 4 contracts in advance, since buyers are unlikely to request services simultaneously. This design improves resource utilization.}. Nevertheless, futures trading with overbooking still encounters significant challenges, as the inherent uncertainty and risk of the future mean that improperly designed long-term contracts can undermine the reputation of the trading market. 

Accordingly, we focus on the last question \textit{Q4. How can the adaptability to the market be achieved?} {To cope with supply-demand fluctuations, we adopt two strategies: \textit{i)} under excess demand, the seller leverages overbooking by selecting a subset of buyers as volunteers\footnote{Volunteers refer to those who temporarily give up their services to purchasing sellers due to resource shortages.} based on pre-signed contracts; \textit{ii)} under excess supply, the seller engages in spot trading via temporary contracts with buyers whose demands remain unmet. However, discrepancies between long-term contracts and real-time market conditions often trigger frequent use of these strategies. This undermines the original purpose of long-term contracts, that is. reducing spot-trading overhead, and may erode participants’ confidence. To address this, we design an adaptive contract update mechanism that monitors market dynamics and participant behaviors in real time, re-evaluating and updating contracts during significant fluctuations to ensure stability, mitigate risks, and preserve market credibility\footnote{Similar practices exist in real systems. For example, cloud data centers use overbooking at the virtualization layer to improve utilization~\cite{virtual layers}. The Singapore Land Transport Authority also requires autonomous vehicle pilots to purchase liability insurance ($\ge$1.5M SGD annually) and submit driving data, where poor safety scores may double premiums or block renewal, thus tying reputation to future service contracts.} }

{Motivated by the above, this paper proposes overbooking- and hybrid-trading-empowered resource scheduling with smart reputation update (Oh-Trust) for edge networks with uncertain demands and conditions. For futures trading, we develop \textbf{bi}lateral \textbf{n}egotiation-based long-term \textbf{c}ontract \textbf{d}esign with \textbf{o}verbooking (BiN\_CDO), which establishes risk-aware and mutually beneficial contracts between sellers (e.g., edge servers) and buyers (e.g., mobile devices) by analyzing uncertainty statistics. For spot trading, we design two contingencies: \textit{i)} voluntary buyers temporarily waive services under shortages in exchange for contractual compensation, and \textit{ii)} \textbf{bi}lateral \textbf{n}egotiation-based \textbf{t}emporary \textbf{c}ontract \textbf{d}esign (BiN\_TCD) balances unmet demand and surplus supply. To mitigate contract breaches and reputation decline, we further propose \textbf{s}mart \textbf{r}eputation \textbf{u}pdating-based \textbf{co}ntract \textbf{r}enew (SRU\_ConR), which applies deep reinforcement learning to evaluate and update contracts, ensuring market stability and credibility.}

\subsection{Investigation}{
This section conducts an extensive investigation on spot, futures, and hybrid trading mechanisms for resource provisioning \cite{Related spot 1,Related spot 2,Related spot 3,Related spot 4,Related spot 5,Related spot 6,Related futures 1,Related futures 2,Related futures 3} in edge networks. A summary of related studies is provided in Table~1. 

\noindent
$\bullet$ \textbf{Spot trading mode.} 
Efficient resource allocation in mobile edge computing (MEC) networks has been mostly studied by following a spot trading mode. 
\textit{Wang et al.} \cite{Related spot 1} developed resource pricing and allocation mechanisms based on a two-stage Stackelberg differential game in UAV-assisted edge networks. 
\textit{Liu et al.} \cite{Related spot 2} proposed a randomized approach combining second-price and fixed-price auctions to address heterogeneous multi-minded resource allocation. 
In \cite{Related spot 3}, \textit{Zhang et al.} designed a pricing strategy that accounts for the effect of CPU temperature on server longevity while meeting QoS requirements for time-sensitive tasks. 
\textit{Xu et al.} \cite{Related spot 4} presented a blockchain-based mechanism and a double auction algorithm for multi-UAV edge computing systems. 
\textit{Ren et al.} \cite{Related spot 5} introduced a space–time–request trading framework for edge–cloud markets. 
\textit{Shih et al.} \cite{Related spot 6} developed a multi-market trading framework for edge computing. 
Although these studies have made significant contributions, they largely depend on onsite decision-making, which suffers from notable drawbacks such as high overhead and potential trading failures. 

\noindent
$\bullet$ \textbf{Futures trading mode.} 
To mitigate the limitations of spot trading, researchers have begun exploring the futures trading mode. 
\textit{Liwang et al.} \cite{Related futures 1} proposed a futures-based methodology for resource trading in UAV-assisted edge networks. 
\textit{Sheng et al.} \cite{Related futures 2} studied spectrum trading using futures contracts in wireless environments. 
{\textit{Liwang et al.} \cite{Related futures 4} introduced a futures-based resource trading approach in edge computing-enabled internet of vehicles.}
However, implementing futures trading presents challenges, particularly when statistical analysis fails to accurately capture future market conditions. These issues may undermine the credibility of the market and reduce buyers' enthusiasm for participation.

	\begin{table}[t!] 

	{\footnotesize
		\caption{\footnotesize{A summary of related studies}}
		\begin{center}\vspace{-.2cm}
			\setlength{\tabcolsep}{0.5mm}{
				\begin{tabular}{|c|c|c|c|c|c|}
					\hline
					\multirow{2}{*}{\textbf{Literature}} & \multicolumn{2}{c|}{\textbf{Trading mode}} & \multicolumn{3}{c|}{\makecell[c]{\textbf{Innovative} \textbf{attributes}}}\\ \cline{2-6} 
					&\makecell[c]{Spot}&\makecell[c]{Futures}&Overbooking&\makecell[c]{Risk\\Analysis}&\makecell[c]{Adaptive Contract\\ Update}\\ \hline
					\makecell[l]{\makecell[c]{\cite{Related spot 1,Related spot 2,Related spot 3},\\ \cite{Related spot 4,Related spot 5,Related spot 6}}} &$\surd$& &&& \\ \hline
					\makecell[l]{\cite{Related futures 1,Related futures 4}} &&$\surd$& &$\surd$& \\ \hline
					\makecell[l]{\cite{Related futures 2} } &&$\surd$ & & &\\ \hline
					\makecell[l]{\cite{Related futures 3}} &$\surd$&$\surd$&$\surd$&$\surd$& \\ \hline
 \makecell[l]{\cite{Related hybrid 1,Related hybrid 2}} &$\surd$&$\surd$& & & \\ \hline
					our work &$\surd$&$\surd$&$\surd$&$\surd$&$\surd$\\ \hline
			\end{tabular}}\label{Tab 1}
	\end{center}}\vspace{-5mm}
\end{table}

\noindent
$\bullet$ \textbf{Hybrid trading mode.} 
More recently, attention has been directed toward hybrid approaches that combine the advantages of both futures and spot trading. 
\textit{Qi et al.} \cite{Related futures 3} investigated cross-layer pre-matching mechanisms to achieve stable and cost-effective hybrid trading-based resource trading over dynamic cloud–edge networks. 
{\textit{Ren et al.} \cite{Related hybrid 1} proposed a space–time–request trading framework for edge–cloud markets, which integrates futures–spot dual trading with dynamic pricing and auction mechanisms to improve resource efficiency. \textit{Huang et al.} \cite{Related hybrid 2} proposed a hybrid market-based resource allocation mechanism for MEC, enabling servers to maximize utility under both complete and incomplete information scenarios.}

In such designs, long-term contracts in the futures stage secure stable demand provisioning, while spot trading complements unmet or bursty demands in real time. 
This two-stage design seeks to balance long-term stability and short-term flexibility, ensuring robust system performance in dynamic edge networks. 
Nevertheless, effectively coordinating futures and spot markets remains a nontrivial challenge, requiring mitigating their shortcomings while maintaining market reputation and participant incentives. A comparative summary of related studies is presented in Table \ref{Tab 1}, highlighting the key distinctions of our methodology.}
{Our methodology offers several distinctive advantages over existing approaches, as summarized below: 

\noindent
$\bullet$ \textbf{Mechanism design:} Integrating overbooking with a hybrid two-stage approach enhances utilization, reduces failures, and adapts to short-term dynamics.

\noindent
$\bullet$ \textbf{Risk management:} Multiple risk constraints (buyer, seller, volunteer) are embedded in long-term contract negotiation, ensuring robustness under uncertainty.

\noindent
$\bullet$ \textbf{Contract renewal:} A reinforcement learning–based reputation update adaptively renews contracts, preserving market credibility and long-term stability.} 
\begin{figure*}[]
	
	\setlength{\abovecaptionskip}{0 mm}
	\centering
	\includegraphics[width=2\columnwidth]{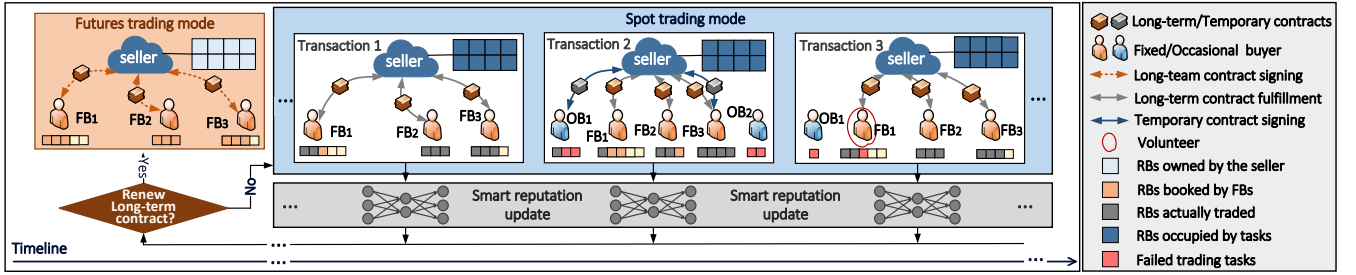}
	\caption{Framework and procedure in terms of a timeline associated with our proposed the Oh-Trust in dynamic edge networks.}\label{Fig 1}
	
\end{figure*}
\subsection{Novelty and Contribution}
To the best of our knowledge, this paper pioneers the integration of overbooking-driven futures and spot trading with intelligent contract renewal in edge networks with dynamic nature. Key contributions are summarized below:

{\noindent
$\bullet$ We are interested in a dynamic edge network with limited resources and unpredictable, heterogeneous demand. To enable a healthy trading market and timely resource sharing between the seller and various buyers, we investigate Oh-Trust, integrating futures and spot trading with a smart contract renewal mechanism.

\noindent
$\bullet$ For futures trading, we propose the BiN\_CDO algorithm, which enables buyers to pre-sign long-term contracts with the seller by determining service prices and default clauses through risk-aware analysis. Notably, incorporating overbooking allows the seller to allocate resources beyond nominal capacity, effectively accommodating dynamic and uncertain demand.}

\noindent
$\bullet$ For spot trading, long-term contracts are performed accordingly. Meanwhile, we also engage two cases during each practical transaction: \textit{i)} the seller may have to determine some voluntary buyers due to resource shortage, while paying them compensations for breaking contracts, and \textit{ii)} we design BiN\_TCD as a complementary plan to enhance resource utilization and market profitability when there exist unmet resource demand and surplus supply.

\noindent
$\bullet$ Given that network dynamics can disrupt the smooth implementation of pre-signed contracts and harm the market's reputation, we develop SRU\_ConR. This incorporates a smart reputation update process that monitors trading performance, assessing whether long-term contracts should be renewed, ensuring adaptability to the continuously evolving market conditions.

\noindent
$\bullet$ We conduct extensive simulations based on a real-world dataset to demonstrate the effectiveness of our proposed Oh-Trust across multiple evaluation metrics.

\section{System Model}
\subsection{Overview}
Our interested service trading market over dynamic edge networks involves two key parties: one seller\footnote{This paper primarily focuses on the single-seller scenario to clearly validate the core mechanisms of Oh-Trust (e.g., overbooking, hybrid trading, and SRU\_ConR). Nevertheless, the framework can be naturally extended to multi-seller markets, where additional challenges such as inter-seller competition and heterogeneous reputations arise. These extensions are outlined as future work of Section 7.}
 (e.g., an edge computing platform) with certain resource supply, virtualized into a resource pool with a certain number of resource blocks (RBs) for analytical simplicity; and multiple buyers which carry different numbers of computing tasks that call for resources from the seller\footnote{Computing resource sharing in Oh-Trust is embodied at multiple layers: 
\textit{i)} sellers slice their computing capacity into allocatable virtual resource units\cite{Related futures 3} (e.g., VMs or container instances), 
\textit{ii)} buyers offload tasks that compete for and share the same server capacity, 
and \textit{iii)} dynamic mechanisms such as overbooking and SRU\_ConR ensure that sharing adapts to demand fluctuations and reputation feedback.}. {Interestingly, we consider two types of buyers: \textit{i)} fixed buyers (FBs) denoted by 	$\bm{\dot{B}}=\left\{\dot{b}_1, ..., \dot{b}_i, ..., \dot{b}_{|\bm{\dot{B}}|}\right\}$, representing frequent resource buyers in the considered market, e.g., residents or fixed devices such as video surveillance cameras in urban monitoring networks and smart meters in residential communities, who request sustaining computing services; and \textit{ii)} occasional buyers (OBs) denoted by $\bm{\ddot{B}}= \left\{\ddot{b}_1, ..., \ddot{b}_j, ..., \ddot{b}_{|\bm{\ddot{B}}|}\right\}$, indicating the buyers who sporadically join in the market, e.g., stochastic passerby, moving vehicles such as autonomous cars, or UAVs temporarily entering the service range during missions, who are interested in temporary computing services when getting access to the seller.} 
More importantly, to capture the dynamic nature of a resource trading market, we focus on several uncertain factors. First, as the resource demand can always be fluctuant, we involve: \textit{i)} ever-changing number of tasks carried by FBs and OBs, and \textit{ii)} uncertain participation of OBs. Then, for the uncertainty related to wireless communications, we consider time-varying wireless channel quality for the link between each buyer and the seller. To facilitate analysis, we also assume that one task can be served by one RB\cite{Related futures 3}.

As one unique aspect in this paper, we integrate futures and spot trading modes into the market\footnote{While both BiN\_CDO (futures trading) and BiN\_TCD (spot trading) use bilateral negotiation, they serve distinct roles: BiN\_CDO establishes long-term contracts offline with risk constraints and overbooking, whereas BiN\_TCD allocates residual resources online via temporary one-shot agreements.}, where futures trading encourages FBs and the seller to sign long-term contracts. Specifically, a long-term contract between a FB $\dot{b}_i\in\bm{\dot{B}}$ and the seller is denoted by \(\dot{\mathbb{C}}_i = \{ \dot{n}_i, \dot{p}, \dot{q}, \tilde{q} \}\), containing three key terms: the number of trading resources ($\dot{n}_i$)\footnote{$\dot{n}_i=0$ indicates that $\dot{b}_i$ will not sign a contract with the seller.}, service price ($\dot{p}$) and default clause ($\dot{q}$ and $\tilde{q}$) for breaking the contract (per single task). Then, our spot trading facilitates the signing of temporary contracts to address unmet resource demands, such as those FBs whose needs are not fulfilled by long-term contracts, as well as OBs, when there remains available resource supply\footnote{To improve scalability, both BiN\_CDO and BiN\_TCD adopt a layered candidate set filtering strategy, where only a reduced set of feasible contracts is iteratively negotiated. Together with the two-stage trading structure (futures + spot), this design effectively reduces the scale of online interactions and computation overhead.}
. A temporary contract $\check{\mathbb{C}}$ primarily includes two key terms: the quantity of trading resources, and the service price $\check{p}$. {Our primary goal is \textit{to establish appropriate contracts between ddiverse buyers and seller, to facilitate a responsive (e.g., time-efficient), cost-effective (e.g., low overhead on decision-making), and healthy (e.g., reputable) trading market over dynamic and uncertain edge networks.}}

Fig. \ref{Fig 1} depicts a schematic of our proposed Oh-Trust in terms of a timeline, where FBs first negotiate long-term contracts with the seller according to the past data/information of historical trading, before future transactions. Then, in each practical transaction, FBs will perform their pre-signed contracts, while others (e.g., OBs and FBs whose needs are not fulfilled by long-term contracts) follow a spot trading mode to sign temporary contracts. In addition, since the dynamic nature of the network may lead to poor performance of long-term contracts in practical transactions, further affecting the market's reputation, our SRU\_ConR is implemented. Specifically, SRU\_ConR functions checking the rationality of long-term contracts, seeing whether they need to be renewed. For example, considering the coexistence of one seller, three FBs (i.e., FB$_1$-FB$_3$), and two OBs (i.e., OB$_1$ and OB$_2$) in Fig. \ref{Fig 1}, FBs first sign long-term contracts with the help of BiN\_CDO. Then, in Transaction 1, the seller's resource supply exactly matches the FB's demand, thanks to the well-designed overbooking strategy. In Transaction 2, the seller may have surplus resources to serve other buyers through BiN\_TCD (e.g., the seller serves OB$_1$ and FB$_3$). {Moreover, in Transaction 3, due to insufficient resources, the seller assigns FB$_1$ as a volunteer, compensating it for the locally executed tasks.}

\subsection{Preliminary Modeling}
We next introduce key parties, i.e., FBs, OBs, and the seller, along with correlative uncertainties.

\noindent
\textbf{Modeling of FBs.} The attribute of each FB $\dot{b}_i \in \bm{\dot{B}}$ is modeled as a 5-tuple $ \left\{ \dot{f}_i, {\dot{g}}_i , {\dot{e}}_i, \dot{r}_i, \dot{\gamma}_i \right\}$, where $\dot{f}_i$ represents its local capability on data processing (e.g., bits/s), ${\dot{g}}_i $ is the transmission power (e.g., watt), and ${\dot{e}}_i$ denotes power consumed by data analysis (e.g., watt). To better describe network dynamics, we consider two uncertain factors: \textit{i)} the uncertain number of tasks in practical transactions, represented as $\dot{r}_i$, and \textit{ii)} the time-varying channel condition for $\dot{b}_i$ to upload its task data, denoted by $\dot{\gamma}_i$, follows a uniform distribution $\dot{\gamma}_i \sim {\bf{U}}(\mu_1, \mu_2)$.

\noindent
\textbf{Modeling of OBs.} Each OB $\ddot{b}_j \in \bm{\ddot{B}}$ can be described as a 5-tuple $\left\{ \ddot{f}_j, {\ddot{g}}_j , {\ddot{e}}_j, \ddot{r}_j, \ddot{\gamma}_j \right\}$, where $\ddot{f}_j$ denotes the local capability on data processing (e.g., bits/s), ${\ddot{g}}_j$ is the transmission power (e.g., watt), ${\ddot{e}}_j$ describes the local computing power (e.g., watt), $\ddot{r}_j$ indicates the amount of required resources, and $\ddot{\gamma}_j$ represents the channel quality regarding the communication link between $\ddot{b}_j$ and the seller. Besides, the number of OBs arriving during each practical transaction can follow a Poisson distribution, denoted by $|\bm{\ddot{B}}| $$\sim$$ \mathbf{P}(\sigma)$, further impacts practical resource demands.

\noindent
\textbf{Modeling of seller.} The seller is modeled as a 3-tuple $ \left\{ \tilde{f}, \tilde{e}, \tilde{r} \right\}$, where $\tilde{f}$ represents its computing capability of data processing (e.g., bits/s), $\tilde{e}$ is the computing power (e.g., watt) on data processing, while $\tilde{r}$ denotes the number of available resources.

\noindent
\textbf{Notations}: For notational and analytical simplicity, we use $\tilde{\square }$, $\dot{\square}$, and $\ddot{\square}$ above certain letters to denote parameters related to the seller, FBs and OBs respectively. Also, we utilize: $f$ to describe the computing capability on data processing, $g$ and $e$ to denote transmission power and computing power, $r$ to represent resource demands and supply, $\gamma$ to evaluate the channel quality of communication links. Besides, notation $\mathbb{C}$ indicates contracts, $p$ and $q$ denote the price and penalty, respectively. 

\section{Proposed BiN\_CDO in Futures Trading}
We start with introducing futures trading of Oh-Trust, which mainly works for FBs since they always stay in the market.
\subsection{Modeling of Utility, Expected Utility and Risk}
We denote the data size of each task as $d$ (e.g., bits). Accordingly, the completion time of a single task of $ \dot{b}_i $ can be calculated by $ \frac{d}{\dot{f}_i} $\footnote{$f_i$ denotes the processing rate in bits/s. Since processing density is measured in cycles/bit, $f_i$ can be equivalently expressed in cycles/s via this conversion. Hence, the ratio $d/f_i$ implicitly captures the task’s processing density without introducing an explicit parameter.}, while that of edge computing (e.g., $ \dot{b}_i $ offloads one task to the seller) is given as
\begin{equation}
	t_{i}^{\text{edge}}=\frac{d}{W\text{log}_2(1+{\dot{g}}_i \dot{\gamma}_i)}+\frac{d}{\tilde{f}},
\end{equation}
where $ W $ denotes the bandwidth of buyers-to-seller communication links, and $ {\dot{g}}_i \dot{\gamma}_i $ indicates the received signal noise ratio (SNR) \cite{Related futures 1}. Thus, for each task of $\dot{b}_i$, we calculate the amount of time that can be saved from enjoying edge computing service, as given by 
\begin{equation}\label{key}{
		\begin{aligned}
			t_{i}^{\text{save}}=\frac{d}{\dot{f}_i}-t^{\text{edge}}_{i},
	\end{aligned}}
\end{equation}
while the saved energy can be calculated as 
\begin{equation}\label{key}{
		\begin{aligned}
			c_{i}^{\text{save}}=\frac{{\dot{e}}_id}{\dot{f}_i}-\frac{d {\dot{g}}_i }{W\text{log}_2(1+{\dot{g}}_i \dot{\gamma}_i)}.
	\end{aligned}}
\end{equation}

Accordingly, we define the unit valuation (per task, e.g., the profit on saving time and energy) that $\dot{b}_i$ can enjoy from obtaining edge service as 
\begin{equation}\label{key}{
		\begin{aligned}
			\dot{v}_i=\omega_1t_{i}^{\text{save}}+\omega_2c_{i}^{\text{save}},
	\end{aligned}}
\end{equation}
where $ \omega_1 $ and $ \omega_2 $ are positive weighting coefficients.
Accordingly, the utility of $\dot{b}_i$ is given as
\begin{equation}
	{\begin{aligned} \small
			&u_i(\dot{\mathbb{C}}_i,\dot{r}_i) = 
			\alpha_i\left(\dot{n}_i\left(\dot{v}_i-\dot{p} \right)\right)\\&+
			(1-\alpha_i)\left(\dot{r}_i\left(\dot{v}_i-\dot{p} \right)-(\dot{n}_i-\dot{r}_i)\dot{q}\right),
	\end{aligned}}
\end{equation}
where $\alpha_i$ is a binary indicator given by (6), describing whether the practical resource demand of $\dot{b}_i$ exceeds the contract.	
\begin{equation}
	{\begin{aligned} \small
			&\alpha_i = 
			\begin{cases} 
				1, & \text{if } \dot{r}_i > \dot{n}_i \\
				0, & \text{if } \dot{r}_i \le \dot{n}_i
			\end{cases}
	\end{aligned}}
\end{equation}
Note that (5) overlooks the situation where a FB with signed contract fails to enjoy edge service, i.e., a FB being selected as a voluntary buyer, due to factors such as insufficient resource supply and raising resource demand. Therefore, let $V$ denote the number of tasks that the seller is expected to process according to pre-signed contracts, which, however, should be executed locally by FBs' themselves, as defined by
\begin{equation}
	{\begin{aligned} \small
			&V = 
			\begin{cases} 
				0, & \text{if } \sum_{\dot{b}_i\in\bm{\dot{B}}}\dot{n}_i^{+} \le \tilde{r} \\
				\sum_{\dot{b}_i\in\bm{\dot{B}}} \dot{n}_i^{+}- \tilde{r}, & \text{if } \sum_{\dot{b}_i\in\bm{\dot{B}}}\dot{n}_i^{+} > \tilde{r}
			\end{cases}
	\end{aligned}}
\end{equation}
where $\dot{n}_i^{+}=\min(\dot{r}_i,\dot{n}_i)$ refers to the quantity of resources that $\dot{b}_i$ can actually utilize in a practical transaction according to the long-term contracts. Since these $V$ tasks can receive compensations from the seller, the sum utility of buyers in $\bm{\dot{B}}$ can be calculated as
\begin{equation}
	{\begin{aligned} \small
			&\dot{U} = \sum_{\dot{b}_i\in\bm{\dot{B}}}u_i(\dot{\mathbb{C}}_i,\dot{r}_i)+V\tilde{q}-\sum_{\dot{b}_i\in\bm{\dot{B}}}u_i(\dot{\mathbb{C}}_i,\mathbbm{v}_i),
	\end{aligned}}
\end{equation}
where $\mathbbm{v}_i$ indicates the number of tasks that $\dot{b}_i$ has to process locally although it has signed a long-term contract with the seller, and we have $\sum_{\dot{b}_i\in\bm{\dot{B}}}\mathbbm{v}_i=V$. Since it is difficult to obtain the exact value of $\dot{U}$ during this time owing to market dynamics, we calculate its expected value as
\begin{equation}\label{key}{\small
		\begin{aligned}
			&\mathbb{E}[\dot{U}]=\sum_{\dot{b}_i\in\bm{\dot{B}}}\mathbb{E}[u_i(\dot{\mathbb{C}}_i,\dot{r}_i)]+\mathbb{E}[V]\tilde{q}-\sum_{\dot{b}_i\in\bm{\dot{B}}}\mathbb{E}[u_i(\dot{\mathbb{C}}_i,\mathbbm{v}_i)]\\&=\mathbb{E}[V]\tilde{q}+\sum_{\dot{b}_i\in\bm{\dot{B}}}\mathbb{E}[\alpha_i]\left((\dot{n}_i-\mathbb{E}[\mathbbm{v}_i])\left(\mathbb{E}[\dot{v}_i]-\dot{p} \right)\right)\\&+
			\sum_{\dot{b}_i\in\bm{\dot{B}}}
			(1-\mathbb{E}[\alpha_i])\\&\times\left((\mathbb{E}[\dot{r}_i]-\mathbb{E}[\mathbbm{v}_i])(\mathbb{E}[\dot{v}_i]-\dot{p} )+(\dot{n}_i-\mathbb{E}[\dot{r}_i]-\mathbb{E}[\mathbbm{v}_i])\dot{q}\right)
			,
	\end{aligned}}
\end{equation}
where derivations of $\mathbb{E}[V]$, $\mathbb{E}[\mathbbm{v}_i]$, $\mathbb{E}[\dot{v}_i]$, $\mathbb{E}[\dot{r}_i]$, and $\mathbb{E}[\alpha_i]$ are detailed by Appendix B. Then, for the seller, the energy consumed on processing one single task is given by
\begin{equation}
	\tilde{c}=\omega_3\frac{d\tilde{e}}{\tilde{f}}+c^{\text{H}},
\end{equation}
where $\omega_3$ refers to a positive weighting coefficient, and \( c^{\text{H}} \) represents the unit monetary cost associated with hardware \cite{Related futures 3,hard}. Accordingly, we consider the utility of seller by involving: \textit{i)} payments from FBs minus service cost, \textit{ii)} compensation from FBs who break contracts, \textit{iii)} compensation paid for voluntary FBs, as given by
\begin{equation}{\small
		\begin{aligned}
			&\tilde{U}=\sum_{\dot{b}_i\in\bm{\dot{B}}}\left(\alpha_i\left(\dot{n}_i\left(\dot{p}-\tilde{c} \right)\right)+ (1-\alpha_i)\left((\dot{n}_i-\dot{r}_i)\dot{q}\right)\right)\\&-V\left(\tilde{q}+(\dot{p}-\tilde{c} )+\dot{q}\right),
	\end{aligned}}
\end{equation}
while the expectation of $\tilde{U}	$ can be calculated as
\begin{equation}{\small
		\begin{aligned}
			&\mathbb{E}[\tilde{U}]=-\mathbb{E}[V]\left(\tilde{q}+(\dot{p}-\tilde{c} )+\dot{q}\right)+\\&
			\sum_{\dot{b}_i\in\bm{\dot{B}}}\left(\mathbb{E}[\alpha_i]\left(\dot{n}_i\left(\dot{p}-\tilde{c} \right)\right)+ (1-\mathbb{E}[\alpha_i])\left((\dot{n}_i-\mathbb{E}[\dot{r}_i])\dot{q}\right)\right).
	\end{aligned}}
\end{equation}

Futures trading generally brings a coexistence of benefits and risks, as the uncertainties can impose losses to both parties. Thus, we assess two risks for $ \dot{b}_i \in \bm{\dot{B}} $.
First, a FB $ \dot{b}_i $ is risking an unsatisfying utility (e.g., the value of (5) turns negative) during a practical transaction, as defined by the probability that the utility of $ \dot{b}_i $ falls below $ u^{\text{\text{min}}} $, given by
\begin{equation}\label{key}{\small
		\begin{aligned}
			\dot{R}_1\left( \dot{b}_i,\dot{\mathbb{C}}_i\right) =\text{Pr}\left(u_i(\dot{\mathbb{C}}_i,\dot{r}_i)<u^{\text{\text{min}}}\right) ,	
	\end{aligned}}
\end{equation}
where $ u^{\text{\text{min}}} $ denotes a tolerable positive value approaching to 0.
Then, as resource overbooking is allowed in our designed market, a FB may be selected as a volunteer and thus forced to process its tasks locally, we show this risk as 
\begin{equation}\label{key}{\small
		\begin{aligned}
			\dot{R}_2\left( \dot{b}_i,\dot{\mathbb{C}}_i \right)=\text{Pr}(\mathbbm{v}_i > 0) .	
	\end{aligned}}
\end{equation}

Besides, the seller also faces the risk on obtaining an unsatisfying utility falling below its desired one (denoted by $\tilde{U}^{\max}$) during a practical transaction, which is defined as 
\begin{equation}\label{key}{\small
		\begin{aligned}
			\tilde{R}\left(\dot{\mathbb{C}}_i\right) =\text{Pr}\left(\tilde{U}<\tilde{U}^{\max}\right) , \forall \dot{b}_i \in \bm{\dot{B}}	
	\end{aligned}}
\end{equation}
When designing long-term contracts, it is crucial to manage the above risks within acceptable ranges. Failure to do so may prompt FBs and the seller to favor spot trading instead.

\subsection{Problem Formulation} 
Our futures trading mode focuses on designing long-term contracts that maximize the expected utilities for both buyers and seller while keeping risks within acceptable limits. This challenge is formulated as a multi-objective optimization (MOO) problem, presented by the following $\bm{\mathcal{P}_1}$.
\begin{subequations}
	\begin{align}
		\bm{\mathcal{P}_1}:~~~&\underset{\dot{\mathbb{C}}_i}{\text{arg max }}~\mathbb{E}{[\tilde{U}]},~ \mathbb{E}{[\dot{U}]} \tag{16}\\
		\text{s.t.}~~~
		&\tilde{c}\le \dot{p} \le \mathbb{E}[\dot{v}_i] , \tag{16a}\\
		&\sum_{\dot{b}_i\in\bm{\dot{B}}}\dot{n}_i \le (1+\tau)\tilde{r}\tag{16b}\\
		&\dot{R}_1\left( \dot{b}_i,\dot{\mathbb{C}}_i\right) \leq \rho_1,~ \forall \dot{b}_i \in \bm{\dot{B}},\tag{16c}\\
		&\dot{R}_2\left( \dot{b}_i,\dot{\mathbb{C}}_i \right) \leq \rho_2,~ \forall \dot{b}_i \in \bm{\dot{B}},\tag{16d}\\
		&\tilde	{R}\left(\dot{\mathbb{C}}_i\right)\leq \rho_3,~~ \forall \dot{b}_i \in \bm{\dot{B}}\tag{16e}
	\end{align}
\end{subequations}
where $ \rho_1 $, $ \rho_2 $, and $\rho_3 $ are risk thresholds falling in $ (0, 1]$. In $ \bm{\mathcal{P}_1} $, constraint (16a) ensures that payments from FBs are sufficient to cover the seller's service costs while also guaranteeing that the expected valuation of $\dot{b}_i$ benefits from the seller can cover the corresponding payment. Constraint (16b) allows the resources to be overbooked by $(1+\tau)\tilde{r}$, where $\tau$ denotes the overbooking rate; constraints (16c), (16d), and (16e) aim to manage risks (derivations detailed by Appendix B). Apparently, problem $\bm{\mathcal{P}_1}$ involves two objectives, each calls for addressing multiple contractual terms, generally imposing NP-hardness. Additionally, probabilistic risk constraints add further complexity to the problem. To address this, bilateral negotiation emerges as a viable solution, enabling different parties with conflicting and self-interested goals to reach a final trading agreement.

\subsection{Solution Design}
\setlength{\intextsep}{10pt} 
\setlength{\textfloatsep}{15pt}
\setlength{\floatsep}{10pt}
\begin{algorithm}[t!] 
	{\setstretch{0.4} 
		\small\caption{{Proposed BiN\_CDO algorithm}}
		\LinesNumbered 
		\textbf{Initialization:} $ \dot{p} \leftarrow p^{\text{max}} $, $\dot{q}\leftarrow \dot{q}^{\text{max}}$, $\tilde{q}\leftarrow \tilde{q}^{\text{max}}$, $\bm{RCC}_i\leftarrow \varnothing$, $\bm{RCC}^*\leftarrow \varnothing$,$\forall\dot{b}_i \in \bm{\dot{B}}$\ 

		\While{$ \dot{p} \ge p^{\min} $}{
			\While{$\dot{q} \ge \dot{q}^{\min}$}{
				\While{$\tilde{q} \ge \tilde{q}^{\min}$}{
					\For{$\forall\dot{b}_i \in \bm{\dot{B}}$}{
						$\dot{n}_i \leftarrow \dot{n}_i^{\max} $
						
						\While{$\dot{n}_i \ge \dot{n}_i^{\min}$}{
							
							\If{meeting (16a), (16c), (16d)}{
								$\dot{\mathbb{C}}_i \leftarrow \{ \dot{n}_i , \dot{p} , \dot{q} , \tilde{q} \}$
								
								$\bm{RCC}_i\leftarrow \bm{RCC}_i\cup\dot{\mathbb{C}}_i$ 
								
							}
							$\dot{n}_i\leftarrow \dot{n}_i - 1 $
						}

					}
					$\tilde{q} \leftarrow \tilde{q}-\triangle \tilde{q}$
				}
				$\dot{q} \leftarrow \dot{q}-\triangle \dot{q}$
			}
			$\dot{p} \leftarrow \dot{p}-\triangle \dot{p}$
		}

		\If{
			$ \bm{RCC}_i \neq \varnothing,~\forall\dot{b}_i \in \bm{\dot{B}} $}{
			\For{$\forall \dot{\mathbb{C}}_i\in \bm{RCC}_i$ }{
				\If{$\dot{\mathbb{C}}_i$ meeting (16a), (16b) and (16e)}{
					$\bm{RCC}^*\leftarrow \bm{RCC}^*\cup \dot{\mathbb{C}}_i $	}
			}

			$\dot{\mathbb{C}}_i^*\leftarrow\underset{ \dot{\mathbb{C}}_i,~\forall\dot{b}_i \in \bm{\dot{B}} }{\text{arg max }} \mathbb{E}[{\tilde{U}}]$, $\dot{\mathbb{C}}_i \in \bm{RCC}^*$

		}
		
		\textbf{Return:} $\mathbb{C}^*_i, ~\forall\dot{b}_i \in \bm{\dot{B}}$}
\end{algorithm}
In this section, we investigate BiN\_CDO to solve $\bm{\mathcal{P}_1}$, with its pseudo-code given in Alg. 1. In particular, BiN\_CDO includes the following key steps: 

\noindent
\textbf{Step 1. Initialization:} At the beginning of Alg. 1\footnote{Although bilateral negotiation can involve iterative updates, the complexity is limited because most contracts are established offline, and the seller evaluates only a finite set of buyer-reported candidates. This keeps the negotiation tractable even in large networks.} , each FB's payment is set by $ \dot{p} = p^{\text{max}} $, default fines $\dot{q}$ and $\tilde{q}$ are set by $\dot{q}^{\text{max}}$ and $\tilde{q}^{\text{max}}$, respectively (line 1, Alg. 1). The amount of resources that $\dot{b}_i$ may purchase is considered as $\dot{n}_i^{\text{max}}$. Specifically, $p^{\text{max}}$, $\dot{q}^{\text{max}} $, $\tilde{q}^{\text{max}}$ and $\dot{n}_i^{\text{max}}$ indicate the upper bound of $\dot{p}$, $\dot{q}$ and $\tilde{q}$, and $\dot{n}_i$, respectively.

\noindent
\textbf{Step 2. Contract term negotiation from FBs' side:} Each FB $\dot{b}_i$ negotiates long-term contract terms, including penalties for breaking contracts $\tilde{q}$ and $\dot{q}$, resource price $\dot{p}$, and the number of trading resources to be purchased $\dot{n}_i$. Specifically, $\dot{b}_i$ evaluates the relevant terms and records candidate contracts that meet related constraints (e.g., (16a), (16c), (16d)) $\bm{RCC}_i$ (a set that records all the candidate contracts). Once all the FBs have decided their candidate contracts, they report the information of $\bm{RCC}_i$ to the seller.

\noindent
\textbf{Step 3. Contract term negotiation from seller's side:} Upon collecting the set of candidate contracts from FB, the seller analyzes each candidate contract, then selects the one that satisfy constraints (16a), (16b), and (16e) and can maximize its expected utility. Finally, the seller signs long-term contracts with some FBs (i.e., $\dot{n}_i>0$), and determines the payment $\dot{p}$, penalty $\dot{q}, \tilde{q}$, and the number of resources $\dot{n}_i$ purchased by $\dot{b}_i$.

\noindent \textit{Optimality Analysis:} 
This paper offers a novel perspective on risk-aware contract negotiation. For FBs, any contract satisfying constraints (16a), (16c), and (16d) can yield positive long-term utility. Constraint (16a) ensures the expected service value offsets resource costs, while (16c) mitigates the risk of unsatisfactory outcomes. Executing tasks locally without edge services yields zero utility. Thus, candidate contracts meeting these constraints provide strong incentives for FBs. From the seller’s standpoint, expected utility is maximized by selecting the most favorable contracts from the reported set.

{\noindent \textit{Complexity Analysis.} 
The computational complexity of BiN\_CDO is on the order of $\mathcal{O}(L_p L_q L_{\tilde{q}} L_n \cdot |\bm{\dot{B}}|)$, where $L_p$, $L_q$, $L_{\tilde{q}}$, and $L_n$ denote discretization steps of contract variables (price $\dot{p}$, penalties $\dot{q}$ and $\tilde{q}$, and resource quantity $\dot{n}$), and $|\bm{\dot{B}}|$ is the number of FBs. Since BiN\_CDO is executed offline, the associated computational burden is concentrated in initialization and remains tractable. Detailed derivation is provided in Appendix~C.}

\section{Proposed BiN\_TCD in Spot Trading}
As market dynamics may prevent the smooth implementation of long-term contracts, while OBs also bring additional resource demands, we encounter two cases in each practical transaction: \textit{Case 1} confronts a resource shortage, where the total resource demand specified in long-term contracts exceeds resource supply. In this case, the seller will randomly select some FBs as voluntary buyers to give up apart of their committed tasks. Then, in \textit{Case 2}, the seller has surplus resources after fulfilling all the long-term contracts, where those buyers with unmet resource demands (e.g., FBs with \(\alpha_i\)=1, and participant OBs), we introduce BiN\_TCD algorithm as a supplementary, helping with enhancing the utilities of both parties. Apparently, the above two cases cannot occur simultaneously, for which we consider random selection for volunteers in the first case, to ensure the fairness among FBs. Then, in the following, we discuss the second case. 

As Case 2 can involve both FBs and OBs, we rewrite the set of buyers that may participate during a practical transaction as $\bm{\check{B}}=\left\{\check{b}_1,...,\check{b}_k...,\check{b}_{|\bm{\check{B}}|} \right\}$ , including FBs with $\alpha_i=1$ and the OBs coming to the market, i.e., $\bm{\check{B}}\in\left\{\dot{b}_i|\alpha_i=1, \forall \dot{b}_i\in\bm{\dot{B}}\right\}\cup\bm{\ddot{B}}$. Each buyer $\check{b}_k$ is modeled as a 5-tuple $\check{b}_k = \left\{ \check{f}_k, \check{g}_k, \check{e}_k, \check{r}_k, \check{\gamma}_k \right\}$.
Since long-term contracts should be performed in this case and we only care about the remaining resources, we also reconsider the seller with its attribute as a 3-tuple $ \left\{ \tilde{f}, \tilde{e}, \tilde{r}^{\prime} \right\}$, where $\tilde{r}^{\prime}$ denotes the number of available resources left over after addressing all the long-term contracts during the practical transaction, i.e., $\tilde{r}^{\prime}=\tilde{r}-\sum_{\dot{b}_i\in\bm{\dot{B}}} \dot{n}_i^{+}$. Note that other symbols have similar meanings to those in Section II. B, we omit their details here due to space limitation. 

\subsection{Modeling of Utility}
Similar to (2), (3), and (4), we can calculate the unit valuation (per task) of $\check{b}_{k}$ as benefit by edge service as
\begin{equation}\label{key}{\small
		\begin{aligned}
			\check{v}_{k}=&\omega_1\left(\frac{d}{\check{f}_k}-\left(\frac{d}{W\text{log}_2(1+\check{g}_k\check{\gamma}_{k})}+\frac{d}{\check{f}_k}\right)\right)+\\&\omega_2\left(\frac{\check{e}_kd}{\check{f}_k}-\frac{d \check{g}_k}{W\text{log}_2(1+\check{g}_k\check{\gamma}_k)}\right).
	\end{aligned}}\tag{17}
\end{equation}
Accordingly, utility of $ \check{b}_k $ in a practical transaction is given by
\begin{equation}\tag{18}
	\check{U}=\sum_{\check{b}_k\in\bm{\check{B}}}\check{n}_k\left(\check{v}_{k}-\check{p}\right),
\end{equation}
where $\check{n}_k$ refers to the number of resource purchased by $ \check{b}_k $. Besides, the utility of seller can be calculated as
\begin{equation}\tag{19}
	\tilde{U}^{\prime}=\sum_{\check{b}_k\in\check{\mathbb{C}}_k}\check{r}_k\left(\check{p}-\tilde{c}\right).
\end{equation}
\subsection{Problem Formulation and Solution Design} 
Our designed spot trading mode aims to facilitate temporary contracts between FBs with unmet demands and OBs, as well as the seller, following the remaining resource supply. Different with futures trading, we aim to maximize the practical utilities of both parties, as formulated by the following MOO problem $\bm{\mathcal{P}_2}$.
\begin{subequations}
	\begin{align}
		\bm{\mathcal{P}_2}:~~&\underset{\check{\mathbb{C}}_k}{\text{arg max }} \check{U},~\tilde{U}^{\prime}\tag{20}\\
		\text{s.t.}~~~
		&\tilde{c}\le \check{p} \le \check{v}_k , \forall\check{b}_k\in\bm{\check{B}},\tag{20a}\\
		&\sum_{\check{b}_k\in\bm{\check{B}}}\check{n}_k \le \tilde{r}^{\prime}.\tag{20b}
	\end{align}
\end{subequations}
Constraint (20a) guarantees that buyers’ payments cover the seller’s service costs while ensuring the valuation $\check{b}_k$ offsets the corresponding payment, and constraint (20b) restricts purchased resources from exceeding the available supply $\tilde{r}^{\prime}$. To address the NP-hardness of $\bm{\mathcal{P}_2}$, we design the BiN\_TCD algorithm, with pseudo-code provided in Alg.~2. To align with futures trading rules and maintain fairness, BiN\_TCD adopts bilateral negotiation as its core, following three steps similar to BiN\_CDO. The main distinction lies in leveraging current market/network information to determine acceptable contract terms. Detailed descriptions are omitted due to space limitations.
\begin{algorithm}[t!] 
	{\small
		\setstretch{0.4} 
		\caption{{Proposed BiN\_TCD algorithm}}
		\LinesNumbered 
		\textbf{Initialization:} $ \check{p} \leftarrow p^{\text{max}} $, $\bm{RCC}_k\leftarrow \varnothing$, $~\forall\check{b}_k\in\bm{\check{B}}$, $\bm{RCC}^*\leftarrow \varnothing$\ 

		\While{$ \check{p} \ge p^{\min} $}{
			\For{$\check{b}_k \in \bm{\check{B}}$}{
				$\check{n}_k \leftarrow \check{n}_k^{\text{max}} $
				
				\While{$\check{n}_k \ge \check{n}_k^{\min}$}{
					
					\If{meeting (20a)}{
						$\check{\mathbb{C}}_k \leftarrow \{ \check{n}_k , \check{p} \}$
						
						$\bm{RCC}_k\leftarrow \bm{RCC}_k\cup\check{\mathbb{C}}_k$ 
						
					}
					$\check{n}_k\leftarrow \check{n}_k - 1 $
				}
				
			}
			$\check{p} \leftarrow \check{p}-\triangle \check{p}$
		}

		\If{
			$ \bm{RCC}_k \neq \varnothing,~\forall\check{b}_k\in\bm{\check{B}} $}{
			\For{$\forall \check{\mathbb{C}}_k\in \bm{RCC}_k$ }{
				\If{$\check{\mathbb{C}}_k$ meeting (20a) and (20b)}{
					$\bm{RCC}^*\leftarrow \bm{RCC}^*\cup \check{\mathbb{C}}_k $	}
			}

			$\check{\mathbb{C}}_k^{*}\leftarrow\underset{ \check{\mathbb{C}}_k,~\forall\check{b}_k\in\bm{\check{B}} }{\text{arg max }} {\check{U}^{\prime}}$, $\check{\mathbb{C}}_k \in \bm{RCC}^*$

		}
		
		\textbf{Return:} $\check{\mathbb{C}}_k^{*},~\forall\check{b}_k\in\bm{\check{B}}$}
\end{algorithm}

\noindent \textit{Optimality Analysis:} 
For buyers, any contract produced by the BiN\_TCD algorithm that satisfies constraint (20a) guarantees positive utility, since the service valuation offsets the purchase cost. Opting out of edge service forces local task execution, yielding zero utility. Hence, identifying candidate contracts under these constraints provides strong incentives for buyers. From the seller’s perspective, utility is maximized by selecting the most favorable contract from the buyer-reported pool.

{\noindent \textit{Complexity Analysis.} 
The computational complexity of BiN\_TCD is given by $\mathcal{O}(L_{\check{p}} L_{\check{n}} \cdot |\bm{\check{B}}|)$, where $L_{\check{p}}$ and $L_{\check{n}}$ are discretization steps of spot price $\check{p}$ and resource quantity $\check{n}$, and $|\bm{\check{B}}|$ is the number of OBs. Compared with BiN\_CDO, BiN\_TCD entails lower computational complexity, making it better suited for real-time spot trading. The complete derivation and comparison are given in Appendix~C.}


%
%
%

\section{Proposed SRU\_ConR}
Since long-term futures contracts are signed in advance based on historical data, estimation errors in demand or channel quality can lead to inefficient contract utilization and degraded trading performance. In practice, market dynamics may cause the contracted resource volume to exceed the seller’s actual supply, forcing contract breaches and penalty costs. Although affected buyers receive compensation, their dissatisfaction can weaken future market participation.
To enhance adaptability in futures trading, we propose SRU\_ConR, a mechanism that dynamically evaluates market reputation by monitoring fluctuations and tracking seller–buyer behaviors in fulfilling or breaching long-term contracts. It further assesses the validity of pre-signed contracts, with reputation values below a threshold signaling the need for contract renewal. An illustrative example is shown in Fig.~\ref{Fig 1}.

\begin{figure}[t!]
	\centering
	\setlength{\abovecaptionskip}{1 mm}
	\includegraphics[width=1\columnwidth]{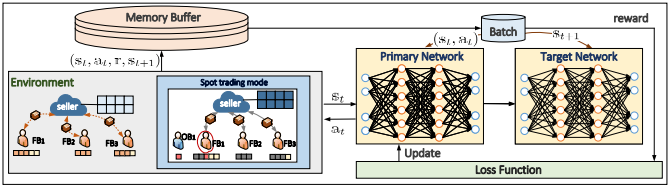}
	\caption{A detailed structure of the SRU\_ConR.}\label{Fig 2}
\end{figure}
\begin{algorithm}[t!] {\small
		\setstretch{0.4} 
		\caption{{Proposed SRU\_ConR}}
		\LinesNumbered 
		
		\textbf{Initialization:} $t=1$, $\theta \leftarrow \text{RandomWeights()}$, $
		\hat{\theta} \leftarrow \theta$
		
		\For{episode = 1, 2, 3, ... }{
			
			$\dot{\mathbb{C}}_i\leftarrow $ Alg.1, $\forall \dot{b}_i\in\bm{\dot{B}}$
			
			\For{$t \le\bm{T} $}{

				Observe state $\mathbbm{s}_{t}$
				
				Select action $\mathbbm{a}_{t}$ using $\varepsilon$-greedy strategy
				
				\If{$\mathbbm{a}_{t}=0$}{$\dot{\mathbb{C}}_i\leftarrow $ Alg.1, $\forall \dot{b}_i\in\bm{\dot{B}}$
					
					$\bm{T}\leftarrow\bm{T}+1$ }

				Get reward $\mathbbm{r}_{t}$ and the next state $\mathbbm{s}_{t+1}^\prime$ 
				
				Store $(\mathbbm{s}_{t}, \mathbbm{a}_{t}, \mathbbm{r}_{t}, \mathbbm{s}^\prime_{t+1})$ into the memory buffer
				
				Sample random minibatch of transitions $(\mathbbm{s}, \mathbbm{a}, \mathbbm{r}, \mathbbm{s}^\prime)$ from memory buffer
				
				\For{ $\forall(\mathbbm{s}_m, \mathbbm{a}_m, \mathbbm{r}_m, \mathbbm{s}_m^\prime)$ in minibatch}{ 
					$Q_m\leftarrow \mathbbm{r}_m + \nu \hat{Q}(\mathbbm{s}^\prime_m, \arg\max_{\mathbbm{a}^\prime} {Q}(\mathbbm{s}^\prime_m, \mathbbm{a}^\prime;{\theta});\hat{\theta})$
					
					Perform gradient descent with loss:
					$L(\theta) = \|Q_m - Q(\mathbbm{s}_m, \mathbbm{a}_m; \theta)\|^2$ \\
					Update \( \theta \) using gradient descent
					
					Update target network parameters:
					$\quad \hat{\theta} \leftarrow \mu \theta + (1 - \mu) \hat{\theta}$
				}

				$t\leftarrow t+1$	
				
			}Continue till function of reward converges
	}}

\end{algorithm}

In particular, SRU\_ConR relies on reinforcement learning which serves as an effective tool, aiming to develop a policy that maximizes long-term rewards (e.g., the long-term performance of the resource trading market). Specifically, the process of our proposed SRU\_ConR can be modeled as markov decision process (MDP) to implement timely updates to long-term contracts. The considered MDP is represented by a 5-tuple $(\mathbb{S}, \mathbb{A}, \nu,\mathbb{P}, \mathbb{R})$\cite{RL,DQN}, where $\mathbb{S}$ is a finite set of states, each element $\mathbbm{s}_t \in \mathbb{S}$ denotes the state at timeslot $t$ (namely, a practical
transaction in our model); The action space is denoted by $\mathbb{A}$, and $\mathbbm{a}_t$ represents an action taken in state $\mathbbm{s}_t$ at timeslot $t$, i.e., $\mathbbm{a}_t \in \mathbb{A}$. Also, $\nu \in [0, 1]$ denotes the discount factor, determining the weight of future rewards during the decision-making process; $\mathbb{P}$ is a Markovian transition model, denoted as $\mathbb{P}(\mathbbm{s}_{t+1}|\mathbbm{s}_{t}, \mathbbm{a}_t)$, representing the probability of transitioning from state $\mathbbm{s}_{t}$ to state $\mathbbm{s}_{t+1}$ when taking an action; while $\mathbb{R}$ describes the reward distribution, denoted by $\mathbb{P}(\mathbbm{r}_t|\mathbbm{s}_t, \mathbbm{a}_t)$, which gives the immediate reward $\mathbbm{r}_t \in \mathbb{R}$ after action $\mathbbm{a}_t$ has been taken in state $\mathbbm{s}_{t}$ at timeslot $t$. Details of state, action, and reward function under the considered MDP framework are given below:

\noindent
$\bullet$ \textit{State:} the state is composed of four components in a practical transaction $t$: \textit{i)} the practical resource demand of FBs (e.g., number of tasks carried by FBs); \textit{ii)} overall utilities of FBs and seller; \textit{iii)} number of interaction between buyers and seller; \textit{iv)} the reputation value $Rep_t$, which combines two parts: the number of tasks successfully completed by following long-term contracts (denoted as $N^{+}_t$) and the number of tasks defaulted (denoted as $N^{-}_t$), calculated as \begin{equation}\tag{21}
	Rep_t=\omega_4N^{+}_t-\omega_5N^{-}_t,
\end{equation} where $\omega_4$ and $\omega_5$ are positive weighting coefficients.

\noindent
$\bullet$ \textit{Action:} the action considers two main factors: \textit{i)} continue with the current long-term contracts ($\mathbbm{a}_t=1 $), and \textit{ii)} long-term contracts should be renewed ($\mathbbm{a}_t=0 $).

\noindent
$\bullet$ \textit{Reward:} the reward function is defined as
\[
\mathbbm{r}_t(\mathbbm{s}_t, \mathbbm{a}_t)=
\begin{cases}\tag{22}
	Rew_1 &,\text{if}~ \mathbbm{a}_t=0 \\
	\omega_6(\dot{U}+\tilde{U})+\omega_7Rep_t &,\text{if}~ \mathbbm{a}_t=1
\end{cases}
\]
where $Rew_1$ represents a negative value incurred when the action of contract renew is chosen. Also, $\omega_6$ and $\omega_7$ are positive weighting coefficients.
{As shown in Fig. \ref{Fig 2}, we utilize double deep Q-networks (DDQN)\cite{DDQN} to implement the renew decision of long-term contracts to better handle the dynamic market with its pseudo-code in Alg. 3}\footnote{Note that the SRU\_ConR module is trained entirely offline, and only lightweight inference is performed online after each transaction to update reputations and decide contract renewal. This design avoids adding significant latency or computational burden to edge devices.}.

\noindent
\textbf{Step 1. Initialization:} The index of practical transactions is initialized by \( t=1 \). Parameter $\theta$ of the primary network (\( Q \)) is initialized with random weights, while the parameter $\hat{\theta}$ of target network (\( \hat{Q} \)) is directly set by \( \theta \) (line 1).

\noindent
\textbf{Step 2. Long-term contracts:} Before practical transactions, FBs and the seller employ BiN\_CDO to determine risk-aware and mutually beneficial long-term contracts by analyzing the statistics of uncertain factors (line 3).

\noindent
\textbf{Step 3. Market reputation analysis and contract updates in spot trading:} In a practical transaction \( t \), each FB \( \dot{b}_i \) and the seller will perform long-term contract \( \dot{\mathbb{C}}_i \), with a possibility of transaction failure due to market dynamics. Given the current market conditions (reputation value \( Rep_t \) from (21), dynamic resource demand \( \dot{r}_i \), etc.), SRU\_ConR uses the $\varepsilon$-greedy strategy\cite{RL,RL2} to decide whether long-term contracts should be renewed. If the decision is made to update long-term contracts (\( \mathbbm{a}_t=0 \)), FBs and the seller will renegotiate new contracts by using BiN\_CDO, incorporating the latest historical data on uncertain factors (e.g., actual resource demands recorded during previous transactions, see line 8). It is worth noting that since BiN\_CDO operates within the futures trading framework, this contract renew operation ($t$) is not considered as a practical transaction, thus we set \( \bm{T}=\bm{T}+1 \) (line 9). Conversely, if the decision is to maintain the current contracts (\( \mathbbm{a}_t=1 \)), the existing contracts remain unchanged and valid.

\noindent
\textbf{Step 4. DDQN model training:} Based on the action determined by our designed DDQN, the corresponding reward can be obtained according to (22), while the state is updated to \( \mathbbm{s}_{t+1} \) (lines 6-8). Next, \( (\mathbbm{s}, \mathbbm{a}, \mathbbm{r}, \mathbbm{s}^\prime) \) is stored in the memory buffer, and a random minibatch of transitions can be sampled from it. For each transition in the minibatch, the target Q-value \( Q_m \) is computed via utilizing the target network \( \hat{Q} \) (line 14), where $\mathbbm{a}^\prime$ represents the optimal action chosen by the primary Q network in the next state denoted by \( \mathbbm{s}^\prime_m \). We then perform gradient descent based on the difference between the target Q-value \( Q_m\) and the primary network \( Q \) (line 15), aiming to minimize the loss function \( L(\theta) \) and update the primary network parameter \( \theta \)\cite{DDQN}. Consequently, the target Q-network \( \hat{Q} \) can be updated by adjusting the target network parameter \( \hat{\theta} \) towards the primary network parameter \( \theta \) (line 17). This process will be repeated until \( t \) reaches \( \bm{T} \) (lines 4-18).

\noindent
\textbf{Step 5. Repetition:} Steps 1-4 are repeated until the value of the reward converges (line 19).

By following the aforementioned steps, the market can intelligently update the reputation associated with long-term contracts and determine whether they should be renewed to better adapt to market dynamics. { Besides, we provide extended discussions on three related aspects in Appendix~D: centralized versus voting-based renewal, global versus individual reputation modeling, and reinforcement learning versus static optimization.}

\section{Evaluation}
We conduct comprehensive evaluations to verify the effectiveness of our proposed Oh-Trust\footnote{To facilitate reproducibility, the public repository at https://github.com/Houyi-Qi/Oh-Trust provides: 
\textit{i)} core module implementations (BiN\_CDO, BiN\_TCD, SRU\_ConR); 
\textit{ii)} baseline implementations; and 
\textit{iii)} data preprocessing scripts for real-world dataset.}
, which are carried out via Python 3.9 with 13th Gen Intel Core i9-13900K*32 and NVIDIA GeForce RTX 4080. 
\subsection{Simulation Settings}
To better emulate a dynamic edge network, we utilize the real-world dataset of Chicago taxi trips\cite{dataset}, which records taxi rides in Chicago from 2013 to 2016 with 77 community areas. Our simulation is interested in the $77^{\text{th}}$ community region with 271259 data points.
To capture the inherent randomness of buyers' demand (e.g., $\dot{r}_i$ and $\ddot{r}_j$), we consider the number of passenger served by a randomly selected taxi in a single day, to estimate the number of tasks carried by a buyer during each practical transaction. Other key parameters are set as follows \cite{Related futures 1,Related futures 3}: $\dot{f}_i=\ddot{f}_j\in[1, 1.5]\times 10^9/600$ bits/s, $ \tilde{f}=1.5\times10^{12}/600 $ bits/s, $\dot{g}_i=\ddot{g}_j\in[500, 550]$ mW, ${\dot{e}}_i= {\ddot{e}}_j\in[450, 500]$ mW, $\tilde{e}=700$ mW, $\mu_1 = 100, \mu_2 = 400$ (i.e., the received SNR thus falls
within $[17, 23]$ dB roughly), $W=6$ MHz, $D=1.5$ Mb, $\tau=0.1$, $p^{\text{max}}=10$, $\dot{q}^{\text{max}}=\tilde{q}^{\text{max}}=5 $, $\sigma=20$, $\tilde{U}^{\max}=2400$, $\dot{q}^{\text{min}}=\tilde{q}^{\text{min}}=1$, $ \rho_1 = \rho_2 =\rho_3 =0.3 $. More importantly, to capture ever-changing resource demands, we randomly select the number of passengers served by a specific taxi over a month (e.g., 30 days) from the dataset as historical data for \( \dot{r}_i \). We then use this data to establish \( \dot{n}_i^{\text{max}} \) and \( \dot{n}_i^{\text{min}} \). Note that \( \dot{r}_i \) in each practical transaction will be recorded as the historical data, and utilized for the proper update of long-term contracts, further capturing the market dynamics.
Without loss of generality, we are inspired by the Monte Carlo method and conduct 1000 simulations for each figure in this section. Namely, values in each figure are averaged. 
\subsection{Benchmark Methods and Evaluation Metrics}
To conduct better evaluations, we involve comparable benchmark methods from diverse perspectives:

\noindent $\bullet$ \textbf{Conventional spot trading method (ConSpot)} \cite{Related spot 3}, which performs a pure spot trading mode.

\noindent $\bullet$ \textbf{Conventional futures trading method (ConFutures)} \cite{Related futures 1}, which implements a pure futures trading mode, without considering the renewal of contracts.

\noindent $\bullet$ \textbf{Hybrid futures and spot trading method (HybridFS)} \cite{Related futures 3}, which is similar to our proposed ``Oh-Trust'', where the key difference refers to that ``HybridFS'' does not allow adaptive and renewable contracts. 

\noindent $\bullet$ \textbf{Random method (Random)} \cite{baseline 3}, in which seller and buyers randomly determine the amount of trading resources and the corresponding prices, serving as a comparative in describing trade-off between time efficiency and trading performance.

More importantly, we also focus on crucial performance metrics detailed below: 

\noindent $\bullet$ \textbf{Utility of buyers and seller:} The overall practical utilities of buyers and seller.

\noindent $\bullet$ \textbf{The proportion of transactions that meet the seller’s desired utility (PoTSU):} PoTSU represents the proportion of transactions, out of 1000 ones, where the seller's actual received utility meets its desired utility.

\noindent $\bullet$ \textbf{The overall value of reputation of the market (VoRM):} VoRM represents the reputation value of a practical transaction conducted under long-term contracts.

\noindent $\bullet$ \textbf{Task completion rate according to long-term contracts (TRLC):} TRLC represents the completion rate of tasks under long-term contracts in a practical transaction, further reflecting their rationality.

\noindent $\bullet$ \textbf{Utilization of Resources (UoR):} UoR represents the utilization of resources of the seller in a practical transaction.

\noindent $\bullet$ \textbf{Running time (RT, ms):} The running time is obtained by Python on verison 3.9, reflecting time efficiency of the market.

\noindent $\bullet$ \textbf{Number of interactions (NI):} Total number of interactions between buyers and seller to reach the final trading decisions, further reflecting the overhead on decision-making.

\noindent $\bullet$ \textbf{Practical task completion time (PTCT, ms):} The PTCT of a task should consider the latency caused by trading decision-making, which can be estimated by the end-to-end delay of communication links between buyers and seller (falls in $[1, 15]$ ms in our simulation \cite{E2E}). Accordingly, PTCT relies on the combined duration of data transmission and processing time, as well as decision-making latency.

\subsection{Performance Evaluations}
\subsubsection{Utility of seller and buyers, and PoTSU}
\begin{figure}[]
	\centering
	\setlength{\abovecaptionskip}{1 mm}
	\includegraphics[width=1\columnwidth]{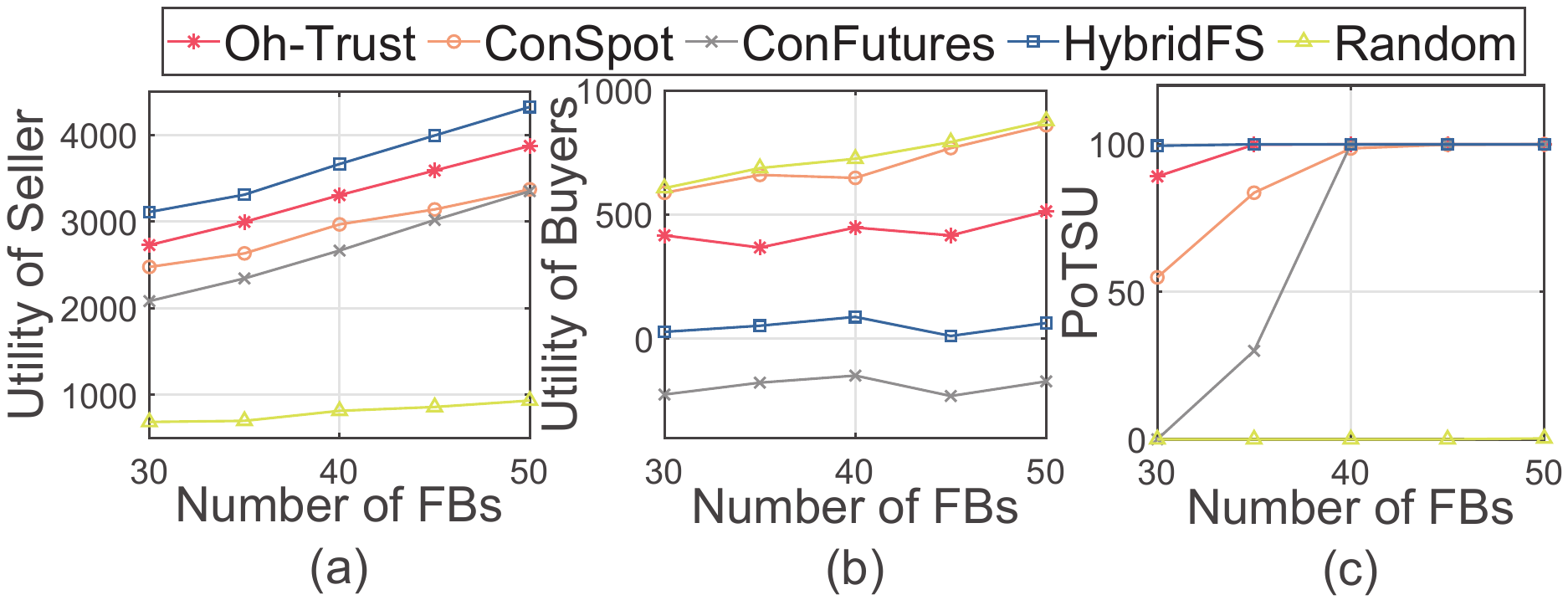}
	\caption{Performance comparisons in terms of different parties upon having $\tilde{r}= 600$: (a) utility of buyers, (b) utility of seller, (c) PoTSU.}\label{Fig U}
\end{figure}
We first examine the average utility of different parties and the PoTSU in Fig.~\ref{Fig U}. In Fig.~\ref{Fig U}(a), as the number of FBs increases, HybridFS yields the highest seller utility since market dynamics often disrupt long-term contracts, leading to FB breaches and penalty payments. Our Oh-Trust surpasses the other benchmarks by integrating multiple trading modes and adaptively renewing contracts. In contrast, ConFutures depends solely on futures trading, suffering performance loss from lacking spot transactions to hedge uncertainty. The Random method performs worst, as random pricing without negotiation fails to safeguard seller utility.
For buyers' utility in Fig.~\ref{Fig U}(b), the Random method performs well by ignoring price negotiations, since buyer and seller utilities are inherently opposed (see Fig.~\ref{Fig U}(a)). Other methods involve multiple negotiation rounds, boosting seller utility but reducing that of buyers (e.g., higher service prices). Our Oh-Trust achieves strong buyer utility, slightly below ConSpot, which optimizes trading based on instantaneous market/network conditions. However, ConSpot incurs heavy decision-making overhead (see Fig.~\ref{Fig NI}), limiting its practicality in dynamic networks. HybridFS and ConFutures perform poorly as they overlook market reputation and lack contract renewal, making them unable to adapt to market dynamics. This results in more FBs breaching contracts and incurring penalties. Fig. \ref{Fig U}(c) also depicts the performance on PoTSU, where the curve of our Oh-trust approaches that of ConFutures and HybridFS and surpasses other methods. {Although Oh-Trust may not always maximize the utility of either side individually, it consistently balances both, and, also together with its superior performance on PoTSU, VoRM, TRLC, and UoR (Figs.~\ref{Fig U} and \ref{Fig Rep}), demonstrates comprehensive advantages over all benchmarks.}


\subsubsection{NI, PTCT and RT}
\begin{figure}[]
	\setlength{\abovecaptionskip}{1 mm}
	\centering
	\includegraphics[width=1\columnwidth]{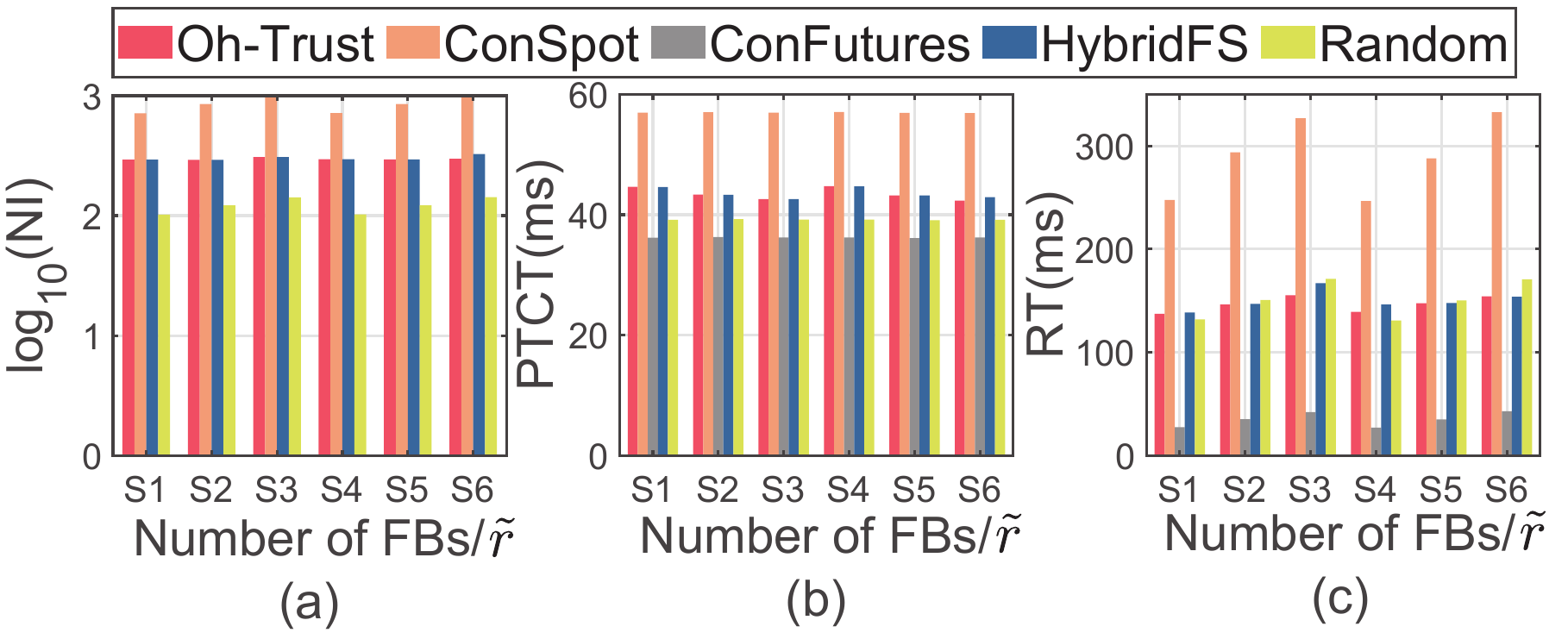}
	\caption{Performance comparisons in terms of: (a) NI, (b) PTCT, and (c) RT under different problem sizes. Specifically, S1-S6 are set as $\{30/500\}$, $\{40/500\}$, $\{50/500\}$, $\{30/600\}$, $\{40/600\}$, and $\{50/600\}$.}\label{Fig NI}
\end{figure}
Since time/energy efficiency plays a significant role in evaluating the performance of a resource trading market with dynamic nature, we next conduct analysis regarding three crucial evaluation indicators, namely, NI (Fig. \ref{Fig NI}(a)), PTCT (Fig. \ref{Fig NI}(b)), and RT (Fig. \ref{Fig NI}(c)), reflecting the overhead incurred by obtaining trading decisions and further the market efficiency. In Fig.~\ref{Fig NI}(a), a logarithmic scale is used to highlight the performance gaps among methods. ConFutures is omitted since it excludes spot trading and thus incurs no decision-making latency, though it performs poorly in participant utilities (see Fig.~\ref{Fig U}). Fig.~\ref{Fig NI}(a) further evaluates NI under varying demand/supply settings (e.g., different numbers of FBs and seller’s supply $\tilde{r}$). Our Oh-Trust and HybridFS outperform ConSpot by leveraging overbooking-enabled futures trading, involving only a small fraction of participants in spot trading and thereby reducing transaction complexity. The Random method yields the lowest NI due to its simplicity (e.g., no price bargaining), but suffers from highly unbalanced utilities (see Fig.~\ref{Fig U}), undermining market stability.
Existing studies often ignore the latency of trading decision-making when evaluating task completion, which is impractical in real networks. We offer a novel perspective on PTCT by incorporating decision-making overhead into actual task completion and compare methods in terms of RT to better capture decision latency. As shown in Figs.~\ref{Fig NI}(b)--\ref{Fig NI}(c), ConFutures performs best in RT, as relying solely on futures trading allows long-term contracts to be executed directly without onsite negotiation. However, this approach struggles under market fluctuations, degrading buyer utility (Fig.~\ref{Fig U}(a)) and TRLC (Fig.~\ref{Fig Rep}(b)). In contrast, the pre-signed contracts in Oh-Trust and HybridFS enable some decisions to be made in advance, streamlining onsite processes and accelerating task completion, as reflected by Oh-Trust’s superior PTCT over ConSpot. {This demonstrates that Oh-Trust achieves a balanced trade-off between utility (Fig.~\ref{Fig U}) and efficiency (Fig.~\ref{Fig NI}), outperforming baselines biased toward either utility maximization or latency minimization.}

\subsubsection{VoRM, TRLC and UoR}
\begin{figure}[]
	\centering
	\setlength{\abovecaptionskip}{1 mm}
	\includegraphics[width=1\columnwidth]{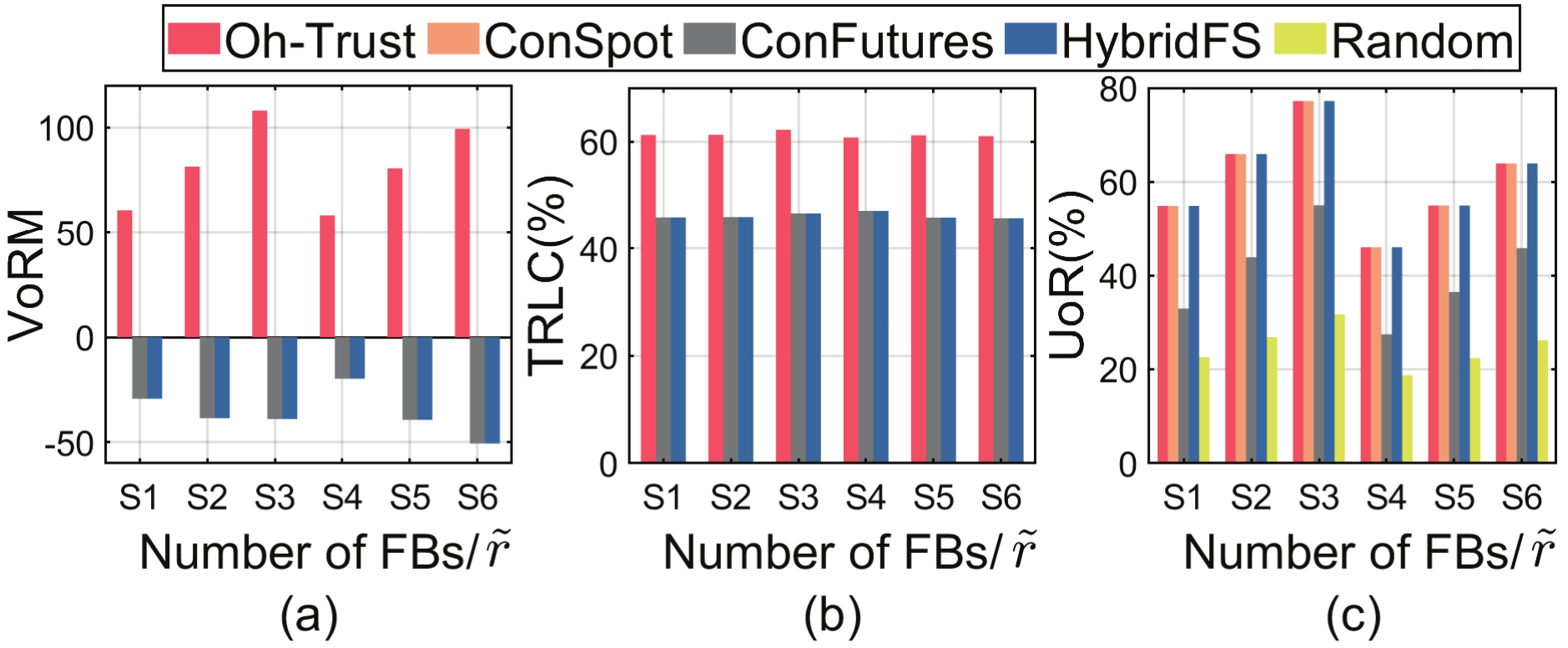}
	\caption{Performance comparisons in terms of: (a) VoRM, (b) TRLC, and (c) UoR under different problem sizes. Specifically, S1-S6 are set as $\{30/500\}$, $\{40/500\}$, $\{50/500\}$, $\{30/600\}$, $\{40/600\}$, and $\{50/600\}$.}\label{Fig Rep}
\end{figure}
Futures trading generally appears to be a coexistance of both profits and risks, bringing the rationality and adaptivity of pre-signed contracts to the front. To handle this issue, intelligently update the reputation of the market and renew the long-term contracts accordingly become one of the most significant considerations in our paper, to against the dynamic and uncertain nature of edge networks. To evaluate the corresponding performance, we conduct Figs. \ref{Fig Rep}(a)-\ref{Fig Rep}(b) regarding VoRM and TRLC.

As can be seen from Fig. \ref{Fig Rep}(a) and Fig. \ref{Fig Rep}(b), our proposed Oh-Trust outperforms ConFutures and HybridFS in terms of VoRM and TRLC, thanks to the well-designed SRU\_ConR. Particularly, Oh-Trust can timely capture the impacts on market reputation (and thus trading performance) brought by implementing long-term contracts in such a ever-changing market, while renewing long-term contracts accordingly. This operation ensures the pre-signed long-term contracts aligns closer with the current network/market conditions during each practical transaction. 

From Fig. \ref{Fig Rep}(c), we can observe that Oh-Trust, HybridFS, and ConSpot methods achieve the best performance on the average value of UoR under different problem scales. This improvement in UoR can be attributed to the implementation of Oh-Trust and HybridFS, which utilize spot trading as a supplementary strategy. By conducting temporary transactions, they effectively address market volatility, enhancing overall UoR. In contrast, the ConFutures method exclusively relies on the futures trading mode, which prevents it from maximizing UoR in dynamic networks. Then, Random method performs the worst because the buyer arbitrarily determines the number of tasks for transactions, leading to inefficient UoR. {Overall, these results confirm that Oh-Trust ensures long-term market stability and adaptability, yielding consistent overall gains even if certain baselines occasionally surpass it in a single evaluation metric.}


\section{Conclusion}
TThis paper presents Oh-Trust, a methodology that unifies the benefits of futures and spot trading in dynamic, uncertain edge networks for timely and cost-efficient resource scheduling. We first develop BiN\_CDO for futures trading, establishing risk-aware, mutually beneficial long-term contracts with overbooking. For spot trading, BiN\_TCD serves as a backup to allocate residual resources and unmet demands. To handle market fluctuations, SRU\_ConR adaptively updates contracts to optimize trading performance. Extensive simulations demonstrate Oh-Trust’s superior performance across multiple metrics. {This work also opens new research directions, including: \textit{i)} extending Oh-Trust to multi-seller edge markets, where fluctuating resource supplies, inter-seller competition, and differentiated service quality would introduce richer and more realistic dynamics; and \textit{ii)} under such competitive scenarios, further refining the characterization of buyer tasks, for example by distinguishing between different types of computing demands or incorporating representative workloads such as deep neural networks (e.g., CNNs). }

\newpage
\clearpage
\appendices
{\section{Key Notations}
Key notations in this paper are summarized in Table \ref{Tab 2}.
\begin{table*}[b!]
\centering
\caption{Key Notations}\label{Tab 2}
{\begin{tabular}{ll}
\toprule
\textbf{Notation} & \textbf{Explanation} \\
\midrule
$\dot{b}_i, \ddot{b}_j, \tilde{s}$ & $i$-th fixed buyer (FB), $j$-th occasional buyer (OB), and the seller. \\
$\bm{\dot{B}}, \bm{\ddot{B}}$ & Sets of FBs and OBs. \\
$\dot{f}_i, \ddot{f}_j, \tilde{f}$ & Computing capability of FB$_i$, OB$_j$, and the seller (bits/s). \\
$\dot{g}_i, \ddot{g}_j$ & Transmission power of FB$_i$ and OB$_j$ (watt). \\
$\dot{e}_i, \ddot{e}_j, \tilde{e}$ & Local/seller computing power (watt). \\
$\dot{r}_i, \ddot{r}_j, \tilde{r}$ & Resource demand of FB$_i$, OB$_j$, and available resources of the seller. \\
$\dot{\gamma}_i, \ddot{\gamma}_j$ & Time-varying wireless channel quality of FB$_i$ and OB$_j$. \\
$d$ & Data size of each task (bits). \\
$t_i^{\text{edge}}, t_i^{\text{save}}$ & Edge execution delay of task $i$; saved time from offloading. \\
$c_i^{\text{save}}, \tilde{c}$ & Saved energy of FB$_i$ by offloading; energy cost per task at the seller. \\
$\dot{v}_i, \check{v}_k$ & Unit valuation (per task) of FB$_i$ or spot buyer $\check{b}_k$. \\
$\dot{u}_i, \dot{U}, \tilde{U}, \check{U}$ & Utility of FB$_i$; total utility of FBs; utility of the seller; total utility of spot buyers. \\
$V$ & Number of tasks forced to be executed locally due to overbooking\\
$\dot{\mathbb{C}}_i = \{\dot{n}_i,\dot{p},\dot{q},\tilde{q}\}$ & 
\makecell[l]{Long-term contract between FB$_i$ and the seller:\\ number of reserved resources, service price, penalty, and compensation.} \\
$\check{\mathbb{C}}, \check{p}$ & Temporary contract in spot trading; spot price. \\
$\dot{n}_i, \check{n}_k$ & Resources reserved for FB$_i$; resources purchased by spot buyer $\check{b}_k$. \\
$\alpha_i$ & Indicator ($1$ if FB$_i$’s demand exceeds contract; $0$ otherwise). \\
$\dot{R}_1, \dot{R}_2, \tilde{R}$ & Risks. \\
$\rho_1, \rho_2, \rho_3$ & Risk thresholds for $\dot{R}_1$, $\dot{R}_2$, and $\tilde{R}$. \\
$\tau$ & Overbooking rate. \\
$W$ & Bandwidth of buyer-to-seller communication links. \\
$\omega_1,\omega_2,\omega_3$ & Weighting coefficients for time saving, energy saving, and seller’s energy cost. \\
$\dot{p}, \check{p}$ & Service prices in futures and spot trading. \\
$\dot{q}, \tilde{q}$ & Penalty and compensation terms in long-term contracts. \\
\bottomrule
\end{tabular}}
\end{table*}}

\section{Derivations Associated with Oh-Trust}
\noindent\textbf{Mathematical expectation of} $\alpha_{i}$. Assume that there are $X_i^{all}$ data points regarding the historical data associated with $\dot{r}_i$, among which, $X_i^1$ data points are greater than $\dot{n}_i$ (e.g., suppose that the historical data of $\dot{r}_i$ has four data points denoted by $\left\{2,3,4,5\right\}$. When considering $\dot{n}_i=3$, we have $X_i^1=2$, since 4 and 5 are lager than 3). The expectation of $\alpha_{i}$ can simply be computed as 
\begin{equation}\tag{23}
	{\small\begin{aligned}
			\mathbb{E}[\alpha_{i}]=1\times\Pr(\dot{r}_i>\dot{n}_i)+0\times \Pr(\dot{r}_i\le \dot{n}_i)=\frac{X_i^1}{X_i^{all}} 
	\end{aligned}}
\end{equation}

\noindent\textbf{Mathematical expectation of} $\dot{v}_i$. $\dot{\gamma}_i$ follows a uniform distribution denoted by $\dot{\gamma}_{i} \sim {\bf{U}}(\mu_1, \mu_2)$. The expectation of $\dot{\gamma}_i$ can simply be computed as $\mathbb{E}[\dot{\gamma}_i] = \frac{\mu_1 + \mu_2}{2}$. According to (1), (2), and (3), we can obtain the value of $\mathbb{E}[\dot{v}_i]$ as
\begin{equation}\tag{24}
	{\small\begin{aligned}
			\mathbb{E}[\dot{v}_i] = \omega_1 &\left\{ 
			\frac{d}{\dot{f}_i} - \left( \frac{d}{\tilde{f}} + \frac{\dot{r}_i}{W \log_2 \left( 1 + {\dot{g}}_i \frac{(\mu_1 + \mu_2)}{2} \right)} \right) \right\} \\
			+ \omega_2&\left\{ 
			\frac{D \dot{e}_i}{\dot{f}_i} - \frac{D {\dot{g}}_i }{W \log_2 \left( 1 + {\dot{g}}_i \frac{(\mu_1 + \mu_2)}{2} \right)} \right\}
	\end{aligned}}
\end{equation}

\noindent\textbf{Mathematical expectation of} $V$ \textbf{and} $\dot{r}_i$. We first use a set $\mathcal{M} = \{M_1, \ldots, M_n, \ldots, M_{|\mathcal{M}|}\}$ to collect all the possible cases of the practical demand from FBs' side, where $M_n = \left\{ \dot{r}_1, \ldots, \dot{r}_i, \ldots, \dot{r}_{|\bm{\dot{B}}|} \right\}$ denotes a vector, which describes the $n^{\text{th}}$ case of FBs' demand in a transaction. For notational simplicity, use a set $\mathcal{M}^{\mathbbm{n}}$ to denote all the cases that $\sum_{\dot{b}_i\in\bm{\dot{B}}}\dot{n}_i^{+} - \tilde{r}=\mathbbm{n}$, where $0<\mathbbm{n}\le\tau\tilde{r}$ (due to overbooked supply), we thus can obtain 
\begin{equation}\tag{25}
	{\small\begin{aligned}
			\Pr\left(V=\mathbbm{n}\right)=\sum_{M_n \in \mathcal{M}^\mathbbm{n}} \prod_{\dot{r}_i \in M_n} \mathbb{E}[\dot{r}_i],
	\end{aligned}}
\end{equation}
where $\mathbb{E}[\dot{r}_i]$ refers to the average value of the historical data of $\dot{r}_i$. To this end, $\mathbb{E}[V]=\sum_{\mathbbm{n}=1}^{\mathbbm{n}=\tau\tilde{r}}\mathbbm{n}\Pr\left(V=\mathbbm{n}\right)$.

\noindent\textbf{Mathematical expectation of} $\mathbbm{v}_i$. We use \( C(*, *) \) to represent the operation on combination. According to (25), we can calculate the probability that $Y$ tasks ($Y\le\mathbbm{n}$) fail to enjoy services when $\dot{b}_i$ is chosen as a voluntary buyer, given by
\begin{equation}\tag{26}
	{\small\begin{aligned}
			&\Pr(\mathbbm{v}_i=Y)=\\&\sum_{\mathbbm{n}=1}^{\mathbbm{n}=\tau\tilde{r}}\Pr\left(V=\mathbbm{n}\right)\frac{C(Y, \dot{n}_i^+)C(\mathbbm{n}-Y, \sum_{\dot{b}_i\in\bm{\dot{B}}}\dot{n}_i^+-\dot{n}_i^+)}{C(\mathbbm{n}, \sum_{\dot{b}_i\in\bm{\dot{B}}}\dot{n}_i^+)}.	
	\end{aligned}}
\end{equation}
Thus, $\mathbb{E}[\mathbbm{v}_i]=\sum_{Y=1}^{Y=\dot{n}_i^+}Y\Pr(\mathbbm{v}_i=Y)$.

\noindent\textbf{Derivations associated with (16d).} From (26), the probability of \(\dot{b}_i\)'s tasks fail to enjoy service when $\dot{b}_i$ is chosen as the voluntary buyer be computed as $\text{Pr}(\mathbbm{v}_i > 0)=\sum_{Y=1}^{Y=\dot{n}_i^+}\Pr(\mathbbm{v}_i=Y)$. Thus, (16d) can be rewritten as:
\begin{equation}\label{key}\tag{27}{\small
		\begin{aligned}
			\dot{R}_2\left( \dot{b}_i,\dot{\mathbb{C}}_i \right)\le \rho_2\Rightarrow\sum_{Y=1}^{Y=\dot{n}_i^+}\Pr(\mathbbm{v}_i=Y)\le \rho_2.
	\end{aligned}}
\end{equation}

\noindent\textbf{Derivations associated with (16c) and (16e).} In optimization problem $\bm{\mathcal{P}_1}$, constraints (16c) and (16e) are in a probabilistic form making their close form non-trivial to be obtained. To resolve this issue, we transform (16c) into a tractable one by exploiting a set of bounding techniques. First, (16c) can be rewritten as
\begin{equation}\tag{28}
	{\small\begin{aligned}
			\dot{R}_{1}(\dot{b}_i, \dot{\mathbb{C}}_i) \leq \rho_1 \Rightarrow \Pr(u_i(\dot{\mathbb{C}}_i, \dot{r}_i) \geq u^{\min}) > 1 - \rho_1. 
	\end{aligned}}
\end{equation}
To obtain a tractable form for (28), we can have upper-bound of its left-hand side by using Markov inequality \cite{2}, as the following (29).
\begin{equation}\tag{29}
	{\small\begin{aligned}
			\Pr(u_i(\dot{\mathbb{C}}_i, \dot{r}_i) \geq u^{\min}) \leq \frac{\mathbb{E}[u_i(\dot{\mathbb{C}}_i, \dot{r}_i)]}{u^{\min}}. 
	\end{aligned}}
\end{equation}
Combining (28) and (29), we then get a tractable form for (16c), as given by
\begin{equation}\tag{30}
	{\small\begin{aligned}
			\frac{\mathbb{E}[u_i(\dot{\mathbb{C}}_i, \dot{r}_i)]}{u^{\min}} > 1 - \rho_1, 
	\end{aligned}}
\end{equation}
where the value of $\mathbb{E}[\tilde{U}]$ is given by (5). Similarly, we can accordingly obtain the tractable form for (16e) as follows
\begin{equation}\tag{31}
	{\small\small\begin{aligned}
			\frac{\mathbb{E}[\tilde{U}]}{\tilde{U}^{\max}} > 1 - \rho_3 .
	\end{aligned}}
\end{equation}

{\section{Complexity Analysis of BiN\_CDO and BiN\_TCD}

Here, we provide detailed analysis of the computational complexity of the two core algorithms within Oh-Trust framework, namely BiN\_CDO (for futures trading) and BiN\_TCD (for spot trading). 

\subsection{Complexity of BiN\_CDO}

The BiN\_CDO algorithm searches over four decision variables: 
\textit{i)} price $\dot{p}$, 
\textit{ii)} penalty $\dot{q}$, 
\textit{iii)} penalty $\tilde{q}$, 
and \textit{iv)} resource quantity $\dot{n}_i$ for each FB. 

\noindent$\bullet$ The outer three loops iterate over $\dot{p}, \dot{q}, \tilde{q}$, respectively. Let the number of possible discrete steps be $L_p$, $L_q$, and $L_{\tilde{q}}$. 

\noindent$\bullet$ For each FB $\dot{b}_i \in \bm{\dot{B}}$ (where $|\bm{\dot{B}}|$ denotes the number of FBs), the inner loop iterates over the resource quantity $\dot{n}_i$ with $L_n$ possible values. 

\noindent$\bullet$ For each candidate contract, the algorithm checks the feasibility of constraints (16a), (16c), (16d), and then collects feasible contracts into the candidate set $\bm{RCC}_i$. 

\noindent$\bullet$ In the final stage, the seller evaluates all candidate contracts $\dot{\mathbb{C}}_i$ and selects the one maximizing its expected utility under (16a), (16b), and (16e). 

Thus, the overall worst-case time complexity of BiN\_CDO can be expressed as:
\begin{equation}\tag{32}
\mathcal{O}\big(L_p \cdot L_q \cdot L_{\tilde{q}} \cdot |\bm{\dot{B}}| \cdot L_n \big).
\end{equation}

Since $L_p, L_q, L_{\tilde{q}}, L_n$ are finite discrete step sizes determined by system parameters, the dominating factor is the number of FBs $|\bm{\dot{B}}|$. Therefore, BiN\_CDO scales linearly with the number of FBs, but the multiplicative factor of the step sizes may increase computational cost in large parameter spaces.

\subsection{Complexity of BiN\_TCD}

BiN\_TCD searches over only two decision variables: 
\textit{i)} price $\check{p}$ and 
\textit{ii)} resource quantity $\check{n}_k$ for each OB. 

\noindent$\bullet$ The outer loop iterates over $\check{p}$ with $L_{\check{p}}$ possible values. 

\noindent$\bullet$ For each OB $\check{b}_k \in \bm{\check{B}}$ (where $|\bm{\check{B}}|$ denotes the number of OBs), the inner loop iterates over $\check{n}_k$ with $L_{\check{n}}$ possible values. 

\noindent$\bullet$ Each candidate contract is checked against (20a) and (20b), and then stored in the candidate set. The seller finally selects the optimal one by maximizing $\check{U}^{\prime}$. 

Thus, the overall worst-case time complexity of BiN\_TCD can be expressed as:
\begin{equation}\tag{33}
\mathcal{O}\big(L_{\check{p}} \cdot |\bm{\check{B}}| \cdot L_{\check{n}} \big).
\end{equation}

Compared with BiN\_CDO, BiN\_TCD involves fewer decision variables and constraint checks, and therefore has a lower computational overhead.}

{\section{Extended Discussions and Clarifications of SRU\_ConR}
To provide additional insights and improve the interpretability of our framework, we include extended discussions on three important aspects of the proposed SRU\_ConR.
\textit{i)} centralized versus voting-based contract renewal, 
\textit{ii)} global versus individual reputation modeling, and 
\textit{iii)} reinforcement learning versus static optimization for contract renewal decisions. 
These clarifications aim to enhance the interpretability, rationality, and broader applicability of our design SRU\_ConR. 

\subsection{Centralized vs. Voting-based Contract Renewal}
The current framework employs a centralized reinforcement learning–driven module (SRU\_ConR) for contract renewal. This choice is motivated by two main considerations: 

\noindent$\bullet$ \textbf{Contract consistency and reputation evaluation.} 
The renewal process depends on the accumulation of historical performance and reputation values of the participants. If each buyer or seller were to cast votes independently, outcomes could be biased by information asymmetry, local utility maximization, or even strategic manipulation. Centralized decision-making enables SRU\_ConR to integrate reputation signals and update contracts from a global perspective, thus ensuring consistency and fairness. 

\noindent$\bullet$ \textbf{System efficiency.} 
Voting-based or consensus-driven approaches would introduce repeated communication rounds, increasing latency and control signaling overhead in resource-constrained edge environments. In contrast, centralized renewal eliminates unnecessary negotiation rounds and improves efficiency. 

Although centralized decision-making fits well with the current design, we also acknowledge that in highly autonomous or cross-domain edge systems, voting- or blockchain-based consensus mechanisms may be more suitable. These approaches can reduce the dependence on a single decision node, improve transparency, and improve robustness. We identify hybrid centralized–decentralized renewal as a promising avenue for future work.

\subsection{Global vs. Individual Reputation Modeling}
The Oh-Trust framework currently employs a \emph{global market reputation} rather than maintaining separate reputations for each participant. This choice is motivated by the following considerations: 

\noindent$\bullet$ \textbf{Complexity and efficiency.} 
Updating and disseminating individual reputations for every participant in each round would significantly increase computational and communication overhead. By contrast, a global reputation serves as an aggregated, system-level signal that quickly reflects the overall reliability of the market. This unified benchmark accelerates bilateral negotiations, reduces redundant communication, and improves fairness. 

\noindent$\bullet$ \textbf{Consistency.} 
Maintaining participant-specific reputations may lead to inconsistent judgments due to asymmetric information, local utility maximization, or even strategic manipulation. A unified global metric mitigates these risks by providing an objective and consistent measure of market trustworthiness. 

\noindent$\bullet$ \textbf{Service quality and stochasticity.} 
Although stochastic models and explicit service-quality indicators (e.g., latency, task completion rate, accuracy) are important for practical deployments, in this work we adopt a deterministic global reputation to maintain tractability and highlight the integration of SRU\_ConR with the hybrid trading framework. Importantly, task service indicators are already embedded in the definition of the reputation value $Rep_{t}$, which jointly accounts for the number of tasks successfully completed and those defaulted. Furthermore, the reward function in (22) explicitly integrates both participants’ utilities ($\dot{U}+\tilde{U}$) and the reputation value $Rep_{t}$, ensuring that economic efficiency and service reliability are simultaneously optimized. Variability is not entirely overlooked: in our simulations, stochastic behaviors are indirectly reflected through task delay distributions and task failure probabilities. 

\noindent\textbf{Future Directions.} 
We envision enriching the reputation mechanism along three lines: 
\textit{i)} stochastic-process–based estimation to capture time-varying uncertainty; 
\textit{ii)} service-quality–driven indicators (e.g., latency, accuracy, completion rate) for finer-grained evaluation; and 
\textit{iii)} hierarchical global–local structures in multi-seller environments, where global reputation provides system-level stability while local reputations capture personalized reliability.

\subsection{Reinforcement Learning vs. Static Optimization for Contract Renewal}
Another key design choice is the use of RL for contract renewal, rather than static optimization. The advantages are as follows: 

\noindent$\bullet$ \textbf{Dynamic adaptability.} 
Static optimization solves for an optimal solution at a given moment, but must be recomputed whenever the environment changes. In contrast, RL continuously updates its policy through interaction with the environment, allowing the renewal mechanism to adapt in real time to fluctuating supply, demand, and reputation. 

\noindent$\bullet$ \textbf{Long-term utility.} 
Contract renewal spans multiple trading rounds. Static optimization often focuses on one-shot objectives, whereas RL (particularly the DDQN used here) optimizes cumulative long-term rewards, incentivizing participants to maintain stable cooperation. 

\noindent$\bullet$ \textbf{Complexity and scalability.} 
The state space of renewal involves multi-dimensional factors (e.g., task completion, risk tolerance, reputation dynamics). Static optimization methods encounter significant challenges in accurately modeling and solving problems with high-dimensional state spaces, while RL leverages function approximation and experience replay to handle complexity more efficiently and scale to larger systems. 

\noindent\textbf{Future Directions.} 
While RL offers strong adaptability and stability, we recognize the value of combining RL with optimization methods. Hybrid approaches may preserve theoretical interpretability while benefiting from RL’s adaptability in dynamic environments.

These extended discussions clarify our design decisions regarding centralized renewal, global reputation, and RL-based adaptability. Together, they reinforce the novelty and feasibility of Oh-Trust while also highlighting promising directions for future research. }

\end{document}